\documentclass[11pt]{article}
\usepackage{amssymb,amsmath,bm,a4,amsthm,array,xy,longtable,graphicx}
\usepackage{rotating}

\newcommand{\eeq}{\end{equation}}
\newcommand{\beq}{\begin{equation}}
\newcommand{\ba}{\begin{array}}
\newcommand{\ea}{\end{array}}
\newcommand{\bea}{\begin{eqnarray}}
\newcommand{\eea}{\end{eqnarray}}
\newcommand{\vev}[1]{\langle #1\rangle}

\newcommand{\preprintno}[1]{\vspace{-2cm}{\normalsize\begin{flushright}#1\end{flushright}}\vspace{1cm}}

\newcommand{\hefo}{${^4}$He}

\newcommand{\lise}{${^7}$Li}
\newcommand{\bese}{${^7}$Be}
\newcommand{\beei}{${^8}$Be}

\newcommand{\mev}{\mbox{MeV}}

\title{\preprintno{HD-THEP-08-18}Unifying cosmological and recent time variations of fundamental couplings}

\author{Thomas Dent,
	Steffen Stern and 
	Christof Wetterich \vspace{0.2cm}\\ 
	{\it Institut f{\" u}r Theoretische Physik, Universit{\" a}t Heidelberg } \\ 
	{\it 16 Philosophenweg, Heidelberg 69120 GERMANY } 
	}
\date{August 2008}

\begin{document}

\maketitle

\begin{abstract}\noindent
A number of positive and null results on the time variation of fundamental constants have been reported. It is difficult to judge whether or not these claims are mutually consistent, since the observable quantities depend on several parameters, namely the coupling strengths and masses of particles. The evolution of these coupling-parameters over cosmological history is also a priori unknown. A direct comparison requires a relation between the couplings. We explore several distinct scenarios based on unification of gauge couplings, providing a representative (though not exhaustive) sample of such relations. For each scenario we obtain a characteristic time dependence and discuss whether a monotonic time evolution is allowed. For all scenarios, some contradictions between different observations appear.
%; hence we expect that one or more observational results are affected by unknown systematic error.
We show how a clear observational determination of non-zero variations would test the dominant mechanism of varying couplings within unified theories. 
\end{abstract}

\section{Introduction}
Any observation of a time variation of ``fundamental constants'' would be a far-reaching discovery. There are various claims for a detection, and many more observations indicating a null variation: criteria for judging their mutual consistency would be useful. We investigate whether a simple scenario exists which can account for several observational claims simultaneously and is consistent with unification of Standard Model (SM) gauge couplings. %as well as with the hypothesis of a cosmological scalar field driving the variations. 
%
%Our aim is to consider whether the available observational data and constraints allow for a simple scenario of (nonzero) variation of fundamental ``constants'' over cosmological time, consistent with gauge unification and with the existence of a cosmological scalar field driving the variation. 
Several observations motivate such a study. First, the claimed deviations from the present value of the fine structure constant $\alpha$, or the proton-to-electron mass ratio $\mu$, observed in quasar absorption systems. Second, the discrepancy between the primordial \lise\ abundance expected from standard nucleosynthesis (BBN) and seen in old halo stars, which may be explained by a variation of ``constants'' within a unified framework \cite{DSW07,CocNunes06}. Third, the theoretical insight that scalar fields cannot be exactly constant over the entire cosmological evolution, and that a possible ``late'' time evolution can play an important role in the dynamics of the expanding Universe. A time variation of couplings arising from the evolution of a ``cosmon'' field in so-called ``quintessence scenarios'' \cite{Wetterich:1987fk,DvaliZ} would link these variations to observables in cosmology \cite{Wetterich:2002ic,Nunes:2003ff}.

Due to the many unknowns of the underlying particle physics models and the partly contradictory present observational situation, a systematic treatment is not easy, and may even seem premature. Nevertheless we consider a first attempt to be useful, in order to discuss strategies that can be used to compare variations of different observables. The power of the proposed method will only become clear if and when future observations present a less ambiguous picture. 

The basic approach in the present paper relates the fractional variations of different fundamental couplings $G_k$, such as the fine structure constant $\alpha$, the proton-electron mass ratio $\mu$ or the ratio of the nucleon mass to the Planck mass, by an assumption of {\it proportionality}, with fixed ``unification coefficients'' $d_k$. The choice of the values of $d_k$ is in turn determined within different scenarios of varying parameters in unified theories (GUTs) where the gauge couplings of the Standard Model converge at a unification scale $M_X$. The assumption of time-independent coefficients $d_k$ covers a large class of possible models for varying couplings. This assumption is, however, not a necessity, and we will describe specific quintessence models where it may not be realized in a forthcoming paper \cite{Part2}.

In Section~\ref{sec:Data} of this paper we review observational determinations of the variation or constancy of couplings, considering five types of methods: early-universe cosmology, astrophysical spectroscopy, nuclear physics in the Earth and the Solar System, gravitational physics, and atomic clock comparisons. In Section~\ref{sec:Unification} we introduce the unification of couplings (Grand Unified Theory, GUT) and determine the implications of unification and supersymmetry (SUSY) for the Standard Model parameters and for the observables we consider. We further define six unified scenarios by considering different possibilities for the variation of the Fermi scale and the superpartner masses. Within each scenario we reduce the various observational results to constraints on the time evolution of a single fundamental coupling, and discuss the mutual consistency of observations.

In Section~\ref{sec:Epochs_and_evolutionFactors} six cosmological epochs are introduced, and the constraints on variation deduced in Section~\ref{sec:Unification} are collected into a set of ``evolution factors'' for each unified scenario. These evolution factors are a measure of the overall size of coupling evolution between a given epoch and the present. We then determine for each scenario to what extent a monotonic variation over time can be consistent with the data. Section~\ref{sec:Conclude} draws some general conclusions.

In a subsequent paper \cite{Part2} we will investigate the scalar field dynamics that could give rise to a small but nonzero variation of couplings. The presence of a cosmologically varying degree of freedom gives rise to important additional effects. It affects gravity on large scales, altering the expansion of the Universe and potentially giving rise to the observed late-time acceleration. Also on local scales, a light field weakly coupled to matter produces long-range forces which are tightly constrained \cite{Will} by Solar System precision tests of gravity and the null results of experiments testing the Weak Equivalence Principle. Combining all these considerations leads to tighter constraints on models but also offers more possibilities to test them.

\section{Data: variations and constraints}
\label{sec:Data}
Here we review and discuss the observational data that we will consider in our effort to obtain a unified picture of time variation of couplings. We summarize the results that are most relevant for our analysis in Table~\ref{tab:CollectionOfMeasurements}.
\begin{sidewaystable}
\center
\begin{tabular}{|c||c|c|c|c|c|c|c|c|c|}
\hline
Method & redshift & $\Delta \ln \alpha $ & $\Delta \ln \mu$ & $\Delta \ln Gm_N^2$ & $\Delta \ln x$ & $\Delta \ln y$ & $\Delta \ln F$ & $\Delta \ln F'$ & $\Delta \ln \lambda_{187}$\\ 
& & $[10^{-6}]$&$[10^{-5}]$&$[10^{-2}]$&$[10^{-5}]$&$[10^{-5}]$&$[10^{-5}]$&$[10^{-4}]$ & $[10^{-2}]$ \\
\hline
\hline
Oklo $\alpha$ \cite{Petrov:2005pu} & 0.14 & $0.00 \pm 0.06$ & & & & & & &\\
\hline
21cm \cite{Murphy:2001nu}       & 0.247 & & & & & $-0.20 \pm 0.44$ & & &\\
Sun \cite{Krauss:1997}          & 0.43 & & & $ 0 \pm 0.72$ & & & & &\\
Heavy/HI, low-z \cite{Tzanavaris:2006uf} & 0.40 & & & & $1.0 \pm 1.7$ & & & &\\
Meteorite \cite{Olive:2003sq}  & 0.44 & & & & & & & & $3.3\pm 3.2$ \\
M$\alpha$ epoch 2 \cite{Murphy:2003mi} & 0.65 & $-2.9 \pm 3.1$ & & & & & & &\\
Ammonia \cite{FlambaumNH3}      & 0.68 & & $0.06 \pm 0.19$ &  & & & & &\\
21cm \cite{Murphy:2001nu}       & 0.685 & & & & & $-0.16 \pm 0.54$ & & &\\
HI / OH \cite{Kanekar:2005xy}   & 0.765 & & & & & & $0.4 \pm 1.1$ & &\\
\hline
Absorption \cite{Fujii:2007zg}  & 1.15 & $-0.1 \pm 1.8$ & &  & & & & &\\
$M\alpha$ epoch 3 \cite{Murphy:2003mi} & 1.47 & $-5.8 \pm 1.3$ & & & & & & &\\
Absorption \cite{Levshakov:2007da} & 1.84 & $5.7 \pm 2.7$ & & & & & & &\\
Heavy/HI, high-z \cite{Tzanavaris:2006uf} & 2.03 & & & & $0.6 \pm 1.9$ & & & &\\
\hline
$H_2$ \cite{Reinhold:2006zn}    & 2.59 & & $2.78 \pm 0.88$ & & & & & &\\
$M\alpha$ epoch 4 \cite{Murphy:2003mi} & 2.84 & $-8.7 \pm 3.7$ & & & & & & &\\
$H_2$ \cite{Reinhold:2006zn}    & 3.02 & & $2.06 \pm 0.79$ &  & & & & &\\
Neutron stars \cite{Thorsett:1996fr} & 3.3 & & & $-0.7 \pm 2.4$  & & & & &\\
CII / CO \cite{Levshakov:2007tn} & 4.69 & & & & & & & $1.4 \pm 1.5$ &\\
CII / CO \cite{Levshakov:2007tn} & 6.42 & & & & & & & $0.1 \pm 1.0$ &\\
\hline
CMB \cite{Martins:2003pe}, \cite{Chan:2007fe} & $10^3$ & $0^{+1\times 10^4}_{-3\times 10^4} $ & & $0^{+7}_{-6}$ & & & & &\\
\hline
\end{tabular}
\caption{Observational $1 \sigma$ bounds on variations. Observables are defined as $\mu\equiv m_p/m_e$, $x\equiv \alpha^2 g_p\mu^{-1}$, $y\equiv \alpha^2g_p$, $F\equiv g_p[\alpha^2 \mu]^{1.57}$, $F'\equiv \alpha^2/\mu$. The given redshift may denote a single measurement, or an averaged value over a certain range: see main text. The two CMB bounds are independent of each other. Our BBN bounds cannot be displayed in this form.} \label{tab:CollectionOfMeasurements}
\end{sidewaystable}

\subsection{Early universe: BBN and CMB}\label{BBN_CMB}
The earliest processes for which Standard Model physics can be tested are BBN ($z \sim 10^{10}$) and CMB ($z \sim 10^3$). Hence they constitute the most far-reaching tests of a possible variation of couplings.

\paragraph{BBN}
The influence of varying constants on BBN has been studied extensively \cite{CocNunes06,LiChu05,Chamoun05,Landau04,MSW,Dmitriev03,ScherrerGN,KnellerLambda,YooScherrer,NollettLopez,
%Ichikawa02,Ivanchik01,
DentF,
%BergstromIR, 
CampbellPrimordialSoup} (see also \cite{Uzanreview} and references therein).
The method developed in \cite{DSW07,MSW} accounts for possible simultaneous variations of different fundamental parameters, while previous studies restricted attention to the variation along one particular direction in parameter space.
% or only single or isolated directions in the space of possible variations. 
%The nice feature about BBN studies is that the result is largely independent of the late-time evolution of the constants of nature. 

Our approach in \cite{DSW07} was first to calculate the leading dependence of BBN abundances on a set of ``nuclear parameters'', comprising elementary particle coupling strengths and masses and nuclear binding energies. In a second step, these nuclear parameters were related to fundamental parameters in particle theory, which allowed us to consider (at linear order) {\em any}\ combination of variations at BBN. Thus, our results are independent of any assumptions about unification.

In an extension of our previous treatment, we include in \ref{app:CMBeta} the possible effect of varying constants at CMB on the input parameter $\eta$ of our BBN procedure. The important parameter is here the variation of $m_N/M_{\rm P}$ at CMB relative to the variation at BBN, where $m_N$ is the nucleon mass and $M_{\rm P}$ is the ``reduced'' Planck mass. We consider two limiting cases. First, when $\Delta (m_N/M_{\rm P})_{|\rm CMB} \ll \Delta (m_N/M_{\rm P})_{|\rm BBN}$: then our previous results hold. In the second case, with $\Delta (m_N/M_{\rm P})_{|\rm CMB} \simeq \Delta (m_N/M_{\rm P})_{|\rm BBN}$, the value of $\eta$ may be significantly rescaled.

Our result for the leading dependence of primordial abundances $Y_a$ on fundamental particle physics parameters $G_k = (G, \alpha, \vev{\phi}, m_e, \delta_q, \hat{m})$ is summarized in Table~\ref{dlnYdlnG}. Here $\delta_q \equiv m_d-m_u$ denotes the light quark mass difference, $\hat{m} \equiv (m_d + m_u)/2$ the average light quark mass, and $\vev{\phi}$ is the expectation value of the Higgs scalar that determines the Fermi scale of the weak interactions. The variation of dimensionful quantities is defined relative to the QCD strong coupling scale $\Lambda_c$. We found that only D, \hefo\ and \lise\ can be used to constrain parameters at BBN. Whilst Table~\ref{dlnYdlnG} only gives linear dependencies, we can account for nonlinearities by running the full BBN code with the appropriate variations \cite{DSW07}.% to obtain bounds on parameters.
\begin{table}
\centering
\begin{tabular}{|c||c|c|c|c|c|}
\hline 
$\partial \ln Y_a/\partial\ln G_k$ &  D   & \hefo & \lise \\ \hline \hline
$G$                                & 0.94 &0.36 & -0.72 \\ \hline 
$\alpha$                           & 3.6  &1.9  &-11    \\ \hline
$\vev{\phi}$                       & 1.6  &2.9  &  1.7  \\ \hline
$m_e$                              & 0.46 &0.40 & -0.17 \\ \hline
$\delta_q$                         &-2.9  &-5.1 & -2.9  \\ \hline
$\hat{m}$                          &17    &-2.7 &-61    \\ \hline
\hline
$\eta$                             &-1.6  & 0.04  &  2.1  \\ \hline
\end{tabular}
\caption{Sensitivity of abundances $Y_i$ to variations of fundamental parameters $G_k$ and the baryon-to-photon ratio $\eta$.}\label{dlnYdlnG}
\end{table}

The current observational and theoretical values for the $Y_a$ are given in Table~\ref{tab:obsBBN}, where $Y_p$ is the helium mass fraction equal to four times the ratio of \hefo\ number density to hydrogen. 
\begin{table}
\center
\begin{tabular}{|c|c|c|}
\hline
Abundance & Observational & Theoretical\\
\hline
 D/H & $(2.8 \pm 0.4) \times 10^{-5}$ & $(2.61 \pm 0.04) \times 10^{-5}$\\
 $Y_p$ & $0.249 \pm 0.009$ & $0.2478 \pm 0.0002$\\
 \lise / H & $(1.5 \pm 0.5) \times 10^{-10}$ & $(4.5 \pm 0.4) \times 10^{-10}$\\
\hline
\end{tabular}
\caption{Current observational and theoretical primordial abundances}
\label{tab:obsBBN}
\end{table}
The uncertainty in the $\eta$ determination, $\eta = (6.20 \pm 0.16) \times 10^{-10}$ (WMAP5 plus BAO and SN, \cite{WMAP5}) yields a further correlated error for the abundances, which can be treated using the method of \cite{Fiorentini:1998fv}. For any given set of fundamental variations we can define
\beq
 \chi^2 \equiv \sum_{i,j} (Y_i - Y_i^{obs}) w_{ij} (Y_j - Y_j^{obs}),
\eeq
with the inverse weight matrix
\beq
 w_{ij} = \left[ \sigma^{2, \eta}_{ij} + \delta_{ij} (\sigma^2_{obs,i} +
 \sigma^2_{th,i}) \right]^{-1},
\eeq
where
\beq
 \sigma^{2, \eta}_{ij} \equiv Y_i Y_j \frac{ \partial \ln Y_i}{\partial \ln \eta}
 \frac{ \partial \ln Y_j}{\partial \ln \eta} 
 \left( \frac{ \Delta \eta}{\eta} \right)^2.
\eeq
We take, as in \cite{DSW07},
\beq
 \frac{\partial \ln ({\rm D/H}, Y_{\rm p}, \mbox{\lise/H})}{\partial \ln \eta} =
 (-1.6,\ 0.04,\ 2.1).
\eeq
The $1(2) \sigma$ error contour is given by $\chi^2/\nu \le 1(4)$ 
%and the $2 \sigma$ error contour by $\chi^2/\nu \le 4$ 
where $\nu$ is the number of degrees of freedom. As the final abundances depend on variations of all fundamental constants, we have to evaluate the variations allowed by BBN for every model separately.

It has been pointed out that the important \beei\ resonance very near the ground states of \bese$+n$ and \lise$+p$ makes the exchange reaction converting \bese\ into \lise\ potentially sensitive to variations in nuclear forces. We give an estimate of this sensitivity in \ref{app:Be8} and show that it is unlikely to be significant for the range of variations that we consider.

In the light of complex astrophysics which may affect the extraction of the primordial \lise\ fraction \cite{Korn06}, we also consider bounding the variations using deuterium and \hefo\ alone. This yields a value consistent with zero for variations at BBN, since these abundances are consistent with standard BBN.%; in this case we cannot claim to solve or ameliorate the ``lithium problem'' by allowing variations.

\paragraph{CMB}
In principle, $\alpha$ and $G$ are bounded by CMB, but there are significant degeneracies with other cosmological parameters \cite{Martins:2003pe,Trotta03}: see also the discussion in Appendix A. Current bounds are
\beq
 0.95<\frac{\alpha_{\rm CMB}}{\alpha_0}<1.02 \qquad (2\sigma). 
\eeq
The CMB anisotropies may also be used to constrain the variation of Newton's constant $G$. The resulting bound depends on the form of the variation of $G$
%\footnote{We drop the suffix unless the context requires it.} 
from the time of CMB decoupling to now. Using a step function one finds \cite{Chan:2007fe,ZahnZaldarriaga02}
\beq 
 0.95 \le \frac{G}{G_0} \le 1.05 \qquad (2 \sigma), 
\eeq
where the 
%redshift $z_s$ of the 
instantaneous change in $G$ may happen at any time between now and CMB decoupling. 
%take any value between $0$ and $\exp(6.8)$, {\em i.e.}\ between now and CMB decoupling, and $z_s$ has been marginalized over. 
Using instead a linear function of the scale factor $a$, the bound is
\beq 
 0.89 \le \frac{G}{G_0} \le 1.13 \qquad (2 \sigma).
\eeq
Note that here, as in most studies of time-dependent $G$, units are implicitly defined such that the elementary particle masses (and thus the mass of gravitating bodies, if gravitational self-energy is neglected) are constant. The relevant bound on {\em dimensionless}\/ parameters concerns $Gm_N^2 \equiv (m_N/M_{\rm P})^2 (8 \pi)^{-1}$.

\subsection{Quasar absorption spectra}
The observation of absorption spectra of distant interstellar clouds allows to probe atomic physics over large time scales. Comparing observed spectra with the spectra observed in the laboratory gives bounds on the possible variation of couplings. Different kinds of spectra (atomic, molecular, \ldots) are sensitive to different parameters, principally $\alpha$ and $\mu \equiv m_p/m_e$.

Atomic spectra are primarily sensitive to $\alpha$. Several groups using various methods of modelling and numerical analysis have published results; we quote here only the latest bounds.
Murphy and collaborators \cite{Murphy:2003mi} studied the spectra of 143 quasar absorption systems over the redshift range $0.2<z_{abs}<4.2$. Their most robust estimate is a weighted mean 
\beq
  \frac{\Delta \alpha}{\alpha} = (-0.57 \pm 0.11) \times 10^{-5}.
\eeq

In discussing unified models in Section~\ref{sec:Epochs_and_evolutionFactors}, we will define various ``epochs'' for the purpose of collating data and comparing them with models over certain ranges of redshift. The 143 data points are then assigned to different epochs: we choose to put boundaries at $z=0.81$ and $z=2.4$, thus we obtain three sub-samples 
\begin{align} 
	z &< 0.81, & N_{sys} &=18, & \vev{z}&= 0.65, & \frac{\Delta\alpha}{\alpha} 
	&= (-0.29 \pm 0.31)\times 10^{-5} \nonumber \\
	0.81 < z &< 2.4 & N_{sys} &=85, & \vev{z}&=1.47, &\frac{\Delta\alpha}{\alpha} 
	&= (-0.58 \pm 0.13)\times 10^{-5} \nonumber \\
	z &> 2.4, & N_{sys} &=40, & \vev{z}&= 2.84, & \frac{\Delta\alpha}{\alpha} 
	&= (-0.87 \pm 0.37)\times 10^{-5}. \label{eq:Murphysplit2}
%	z &> 0.81, & N_{sys} &=125, & \vev{z}&= 1.91, & \frac{\Delta\alpha}{\alpha} 
%	&= (-0.61 \pm 0.12)\times 10^{-5}, \label{eq:Murphysplit2}
\end{align}
Here we have used the ``fiducial sample'' of \cite{Murphy0306},
%including increased error bars for ``high-contrast'' systems
the weighted average has been taken, and we have included \cite{Murphyprivate} the 15 additional samples used in \cite{Murphy:2003mi}. For convenience we will refer to these results as ``M$\alpha$''.

Further results have been obtained by Levshakov {\em et al.}\ \cite{Levshakov:2007da}, and reported in \cite{Fujii:2007zg}:
\begin{align}
 \frac{\Delta \alpha}{\alpha} &= (-0.01 \pm 0.18)\times 10^{-5}, & z_{abs} &= 1.15 \nonumber \\
 \frac{\Delta \alpha}{\alpha} &= (0.57 \pm 0.27)\times 10^{-5}, & z_{abs} &= 1.84.
\end{align}
Previously \cite{Srianand04,Levshakov:2005ab} more stringent null results were claimed, but doubts have been cast \cite{Murphycrit} (see also \cite{Molaro:2007kp}) on the validity of the statistical analysis. We note that the value for $z=1.84$ has an opposite sign of variation to the M$\alpha$ result, though the variation does not have high statistical significance. The observational situation is clearly unsatisfactory.

Vibro-rotational transitions of molecular hydrogen $H_2$ are sensitive to $\mu \equiv m_p/m_e$. From H$_2$ lines of two quasar absorption systems (at $z = 2.59$ and $z=3.02$) a variation is found \cite{Reinhold:2006zn} of
\beq \label{Reinholdmu}
 \frac{\Delta \mu}{\mu} = (2.4 \pm 0.6) \times 10^{-5},
\eeq
taking a weighted average. We will refer to this result as ``R$\mu$'' after Reinhold {\it et al}. The individual systems yield \cite{Reinhold:2006zn}
\begin{align} \label{eq:Reinholdsystems}
 \frac{\Delta \mu}{\mu} &= (2.78 \pm 0.88) \times 10^{-5}, & z_{abs} &= 2.59 \nonumber \\
 \frac{\Delta \mu}{\mu} &= (2.06 \pm 0.79) \times 10^{-5}, & z_{abs} &= 3.02. 
\end{align}
Recently the $z=3.02$ system has been reanalyzed \cite{Wendt08}, with the result that the claimed significance of Eq.~(\ref{eq:Reinholdsystems}) was not reproduced, and the absolute magnitude of the variation is bounded by $|\Delta \mu / \mu| \le 4.9 \times 10^{-5}$ at $2 \sigma$, or
\beq
 |\Delta \mu / \mu| \le 2.5 \times 10^{-5}, \qquad z_{abs} = 3.02 \quad (1 \sigma). 
\eeq
%For comparison with the R$\mu$ result at redshift 2.5--3 we may select the 50 highest-redshift systems of M$\alpha$ which have $\vev{z}=2.72$, with a weighted mean variation of $\Delta \ln \alpha = (-0.85\pm0.33)\times 10^{-5}$. However this value is perfectly consistent with the total high-$z$ sample of Eq.~\ref{eq:Murphysplit2} and the difference is not significant.

The inversion spectrum of ammonia has been used to bound $\mu$ precisely at lower redshift \cite{FlambaumNH3}. Recently the single known NH$_3$ absorber system at cosmological redshift has been analysed \cite{MurphyNH3}, yielding
\beq
 \frac{\Delta \mu}{\mu} = (0.74\pm 0.89) \times 10^{-6}, \qquad z = 0.68.
% old number (0.6\pm 1.9 ) \times 10^{-6}, \qquad z = 0.68.
\eeq
This is a considerably stricter bound but applies at a different epoch. Extrapolation to today with linear time dependence gives $\dot{\mu}/\mu = (1.2 \pm 1.4)\times 10^{-16}\,$y$^{-1}$.

The 21cm HI line and molecular rotation spectra are sensitive to $y \equiv \alpha^2 g_p$, where $g_p$ is the proton g-factor. Bounds on this quantity from \cite{Murphy:2001nu} are
\begin{align}
\frac{\Delta y}{y} &= (-0.20 \pm 0.44) \times 10^{-5}, &z &= 0.247 \nonumber\\
\frac{\Delta y}{y} &= (-0.16 \pm 0.54) \times 10^{-5}, &z &= 0.685.
\end{align}

Further, the comparison of UV heavy element transitions with HI line probes for variations of $x \equiv \alpha^2 g_p \mu^{-1}$  \cite{Tzanavaris:2006uf}: the weighted mean of nine analysed systems yields
\beq
  \frac{\Delta x}{x} = (0.63 \pm 0.99) \times 10^{-5},\quad 0.23<z_{abs}<2.35.
\eeq
However, we note that i) the systems lie in two widely-separated low-redshift ($0.23<z<0.53$) and high-redshift ($1.7<z<2.35$) ranges; and ii) these two sub-samples have completely different scatter, $\chi^2/\nu$ about the mean for the low- and high-redshift systems being $0.33$, and $2.1$, respectively. Hence we consider two samples, with average redshift $z=0.40$ (5 systems) and $z=2.03$ (4 systems). With expanded error bars in the high-redshift sample 
%to account for potential systematic scatter 
(after ``method 3'' of \cite{Tzanavaris:2006uf}) we find
\begin{align} 
 \frac{\Delta x}{x} &= (1.02 \pm 1.68) \times 10^{-5}, & \vev{z} &= 0.40 \nonumber \\
 \frac{\Delta x}{x} &= (0.58 \pm 1.94) \times 10^{-5}, & \vev{z} &= 2.03. 
\end{align}

The comparison of HI and OH lines is sensitive to changes in $F \equiv g_p \left[ \alpha^2 \mu \right]^{1.57}$ \cite{Kanekar:2005xy} and yields
\beq
 \frac{\Delta F}{F} = (0.44 \pm 0.36^{stat} \pm 1.0^{sys}) \times 10^{-5},\quad z = 0.765. 
\eeq
A similar method comparing CII and CO lines has very recently been proposed at high redshift \cite{Levshakov:2007tn} yielding the best bound at redshifts $> 4.5$. The following bounds on $F'\equiv \alpha^2/\mu$ 
%are obtained at the $10^{-4}$ level 
are obtained for two systems:
\begin{align} 
 \frac{\Delta F'}{F'} &= (0.1 \pm 1.0)\times 10^{-4}, & z & = 6.42 \nonumber \\
 \frac{\Delta F'}{F'} &= (1.4\pm 1.5)\times 10^{-4},  & z & = 4.69. 
\end{align}

\subsection{Terrestrial and Solar System nuclear constraints}
\paragraph{Oklo natural reactor}
From modelling nuclear reaction processes which happened in the Oklo mine about two billion years ago ($\Delta t\simeq 1.8\times 10^9\,$y, $z \sim 0.14$ with WMAP5 best fit cosmology) one can in principle bound the variation of $\alpha$ over this period. The determination of $\Delta \ln \alpha$ at the time of the reactions results from considering the possible shift, due to variation of electromagnetic self-energy, in the position of a very low-lying neutron capture resonance of $^{149}$Sm. The analysis of \cite{Petrov:2005pu} gives the bound (taken as 1$\sigma$) 
\beq \label{Okloalpha}
 -5.6\times 10^{-8} < \Delta\alpha/\alpha < 6.6\times 10^{-8}.
\eeq
For a linear time dependence this results in the bound%of $\alpha$ 
\[
%-4 \times 10^{-17} {\rm y}^{-1} < 
 |\dot{\alpha}/\alpha| \leq 3 \times 10^{-17} {\rm y}^{-1}. 
\]
A more recent analysis using different reactor models and consequently different neutron spectra \cite{Gould} found
\[
 -2.4 \times 10^{-8} \leq \Delta \alpha/\alpha \leq 1.1\times 10^{-8}
\]
with an additional nonzero solution (due to the other branch of the resonance peak) at $\Delta \alpha/\alpha \simeq 8\times 10^{-8}$. We will consider the more conservative bound.

Note that these results concern varying $\alpha$ only. If other parameters affecting nuclear forces, in particular light quark masses, are allowed to vary, the interpretation of this bound becomes unclear \cite{Olive:2002tz, Flambaum:2002wq} since it depends on a nuclear resonance of ${^{150}}$Sm whose properties are very difficult to investigate from first principles. In the absence of a resolution to this problem we consider Oklo as applying only to the $\alpha$ variation in each model. In scenarios where several couplings vary simultaneously we do not consider strong cancellations. Nevertheless, we allow for a certain degree of accidental cancellation and therefore multiply the error on the bound Eq.~(\ref{Okloalpha}) by a factor three.
%\footnote{Since only the averaged neutron capture cross-section is experimentally bounded, we cannot be sure {\em a priori}\ that the depletion of ${^{149}}$Sm in the Oklo reactor was due to the same resonance as found today.}

\paragraph{Meteorite dating}
Long-lived $\alpha$- or $\beta$-decay isotopes may be sensitive probes of cosmological variation \cite{Olive:2003sq,Olive:2002tz,SisternaV}. Their (generally) small $Q$-values result from accidental cancellations between different contributions to nuclear binding energy, depending on fundamental couplings in different ways, thus the sensitivity of the decay rate may be enhanced by orders of magnitude.

The best bound concerns the ${^{187}}$Re $\beta$-decay to osmium with $Q_\beta=2.66\,$keV. The decay rate $\lambda_{187}$ is measured at present in the laboratory, and also deduced by isotopic analysis of meteorites formed about the same time as the solar system, $t_{\rm Met}\simeq 4.6\times 10^{9}\,$y ago ($z \simeq 0.44$). More precisely, the ratio $\lambda_{187}/\lambda_{\rm U}$, averaged over the time between formation and the present, is measurable \cite{Olive:2003sq,FujiiMeteorites}, where $\lambda_{\rm U}$ is the rate of some other decay (for example uranium) used to calibrate meteorite ages.
%\[ 
%| \Delta \alpha / \alpha | < (-8 \pm 8) \times 10^{-7},
%\]
The experimental values of $\lambda_{187}$ imply (setting $\lambda_{U}$ to a constant value)
\beq \label{meteoriteintegral}
	t_{\rm Met}^{-1} \int_{-t_{\rm Met}}^{0} \frac{\Delta \lambda_{187}(t)} {\lambda_{187}} 
%\frac{\lambda_{187}(t)- \lambda_{187}(t=0)} {\lambda_{187}(t=0)} 
	\, dt = 0.016\pm 0.016 
 %\equiv \left(\frac{\Delta \lambda_{187}}{\lambda_{187}}\right)_{\rm exp}.
.
\eeq
Since the redshift back to $t_{\rm Met}$ is relatively small, we obtain bounds on recent time variation by assuming a linear evolution up to the present, for which the LHS is $-(t_{\rm Met}/2) \dot{\lambda}_{187}/ \lambda_{187}$ and the fractional rate of change is bounded by 
\beq
	\frac{\dot{\lambda}_{187}}{\lambda_{187}} \simeq (-7.2\pm6.9) \times10^{-12}\,{\rm y}^{-1}.
\eeq
Projected back to $t_{\rm Met}$ this gives the bound $\Delta \ln \lambda_{187} \simeq 0.033\pm0.032$ ($z\simeq 0.44$). This is a conservative bound unless the time variation has recently accelerated (which we consider unlikely), or there are significant oscillatory variations over time.

The decay rate varies as \cite{Olive:2002tz}
\[
	\lambda_{187} \propto G_F^2 Q_\beta^3 m_e^2 \propto \vev{\phi}^{-2} Q_\beta^3 y_e^2 
\]
where $y_e$ is the electron Yukawa coupling.
%(whose variation we set to zero in most scenarios). 
Variation in $Q_\beta$ is determined by the near-cancellation between the nuclear Coulomb self-energy and asymmetry energy of ${^{187}}$Re and ${^{187}}$Os, and the nucleon masses, via
\[
	\Delta {Q_\beta} \simeq 0.77 \Delta {a_A} - 26 \Delta {a_C} + \Delta (\delta_N-m_e), 
\]
where $a_A\simeq 23.7\,$MeV and $a_C\simeq 0.71\,$MeV are the asymmetry and Coulomb terms of the semi-empirical mass formula and $\delta_N\equiv m_n-m_p$ is the nucleon mass splitting. Thus the fractional variation in $\lambda_{187}$ is
\beq \label{Deltalnlambda187nuc}
	\Delta \ln \frac{\lambda_{187}}{m_N} \simeq 2.1\times 10^4\, \Delta \ln \frac{a_A}{m_N} 
	- 2.1\times 10^4\, \Delta \ln \alpha + 880\, \Delta \ln \frac{\delta_N-m_e}{m_N}. 
\eeq
Since the possible dependence of ``control'' decay rates $\lambda_{\rm U}/m_N$ on nuclear or fundamental parameters is much weaker than that of $\lambda_{187}/m_N$, we use this result for the variation of $\lambda_{187}$ in units where $\lambda_{\rm U}$ is constant.%, {\it i.e.}\ $\Delta \ln (\lambda_{187}/\lambda_{\rm U})\simeq \Delta \ln (\lambda_{187}/m_N)$.
\footnote{Variation of $\Lambda_c$ alone would not cause a dominant effect on $\lambda_{187}$. Both the Coulomb and asymmetry terms scale with $\Lambda_c$, thus the effect of varying $\Lambda_c$ is confined to the last term on the RHS of (\ref{Deltalnlambda187nuc}), {\it i.e.}\ varying the ratio of the weak scale to the strong scale.}
Then using relations 
%for the variation of $(\delta_N-m_e)$ and $a_A/m_N$ 
derived in \cite{DentEotvos} 
and considering also the effect of varying $m_s$ on the nucleon mass, 
\begin{align}
	\Delta \ln \frac{\delta_N-m_e}{m_N}&\simeq 2.6\,\Delta \ln \frac{\delta_q}{\Lambda_c}
	-0.65\,\Delta \ln \frac{m_e}{\Lambda_c} -0.97\, \Delta \ln \alpha 
%	-0.048 \Delta \ln \frac{\hat{m}}{\Lambda_c} 
	-0.12 \Delta \ln \frac{m_s}{\Lambda_c}, 
	\label{DeltalnQnmN} \\
	\Delta \ln \frac{a_A}{m_N} &\approx -0.9\, \Delta \ln \frac{\hat{m}}{\Lambda_c},
	\label{DeltalnaAmN}
\end{align}
we find the decay rate dependence to be
\beq
	\Delta \ln \frac{\lambda_{187}}{m_N} \simeq 
	-2.2\times 10^4 \Delta \ln \alpha - 1.9\times 10^4 \Delta \ln \frac{\hat{m}} 
	{\Lambda_c} + 2300\, \Delta \ln \frac{\delta_q}{\Lambda_c}
	-580\, \Delta \ln \frac{m_e}{\Lambda_c} .
\eeq

\subsection{Bounds on the variation of the gravitational constant}
Variations of Newton's constant have been studied in the solar system and in astrophysical effects. Whilst all references give bounds exclusively on a potential variation of $G$, one should note that besides $G$ also nuclear parameters (neutron / proton masses and parameters of nuclear forces) can vary, which would in general add degeneracies and make the results less stringent. It has generally been assumed that particle masses are constant, thus the resulting bounds actually constrain variation of $Gm_N^2 \propto (m_N/M_{\rm P})^2$.

In the solar system, changes of $G$ induce changes in the orbits of planets. Range measurements to Mars from 1976 to 1982 can be used to obtain \cite{Hellings83}
\beq
 \dot{G}/G = (2 \pm 4) \times 10^{-12} \mbox{y}^{-1}. 
\eeq
%which apply at the present epoch. 
Lunar laser ranging from 
%
%1969 to 1990 yields (see Mueller, Schneider, Soffel, Ruder, Astrophys. J. 382 (1991) %L101)
%\[\dot{G}/G = 0.1 \pm 10.4 \times 10^{-12} \mbox{yr}^{-1}$$
%A slightly different dataset of Lunar laser ranging from 
1970 to 2004 yields \cite{Williams04}
\beq 
 \dot{G}/G = (4 \pm 9) \times 10^{-13} \mbox{y}^{-1}. 
\eeq
The stability of the orbital period of the binary pulsar PSR 1913+16 \cite{Damour88} may be used to deduce 
\beq
 \dot{G}/G = (1.0 \pm 2.3) \times 10^{-11} \mbox{y}^{-1}. 
\eeq
Recent observational advances may improve such bounds considerably, with a limit of 
\beq 
 \dot{G}/G = (−0.5 \pm 2.6) \times 10^{-12} \mbox{y}^{-1}
\eeq 
from PSR J0437−4715 quoted in the preprint \cite{Deller:2008jx}. All these results apply at the present epoch $z=0$.

A bound on the behaviour of $G$ over the lifetime of the Sun (approximately $4.5\times 10^{9}$y, $z = 0.43$) was found by Krauss {\it et al.} \cite{Krauss:1997} by considering the effect of the resulting discrepancy in the helium/hydrogen fraction on p-mode oscillation spectra. The claimed constraint is
\begin{align}
 |\dot{G}/G| &\leq 1.6 \times 10^{-12}\,{\rm y}^{-1} \nonumber \\
 |\Delta \ln G| &\leq  7.2 \times 10^{-3} \qquad z=0.43,
\end{align}
where the assumed form of variation is a power law in time since the Big Bang, which may be approximated over the last few billion years as a linear dependence. For models with significantly nonlinear time dependence the bound may be reevaluated: since the bound arises from the accumulated effect of hydrogen burning since the birth of the Sun, it may be expressed as an integral of the variation over the Sun's lifetime analogous to Eq.~(\ref{meteoriteintegral}).

The properties of compact objects such as white dwarfs and neutron stars (NS) have been used to bound possible variations of $G$: see for example \cite{Jofre:2006ug,Benvenuto:2004bs}. The strongest bound not relying on speculative physical effects arises from comparing the masses of young and old neutron stars in binary systems \cite{Thorsett:1996fr}: if one member of the binary is a pulsar, precision timing can be used to determine the masses. The mass of neutron stars at formation is determined to first approximation by the Chandrasekhar mass 
\beq
 M_{\rm Ch} \simeq \frac{ 1%\hbar^{3/2} c^{3/2}}
}{G^{3/2} m_n^2} 
\eeq
where $m_n$ is the neutron mass. This may be reexpressed in terms of the baryon number of the star $n_B\propto M_{\rm Ch}/m_n \propto (Gm_n^2)^{-3/2},$ which is expected to be conserved up to small corrections from matter accreting onto it. Thus the relative masses of NS measured at the same epoch probes the fractional variation of $Gm_n^2$ between their epochs of formation. Given that the oldest neutron stars are up to 12 Gy old, $z \sim 3.3$, the variation of the average NS mass $\mu_n$ is found to be 
$ \dot{\mu}_n = -1.2 \pm 4.0 (8.5) \times 10^{-12} M_{\odot}\, \mbox{y}^{-1} $
at 60\% (95\%) confidence level. 
% We will interpret it as bounding variation of $Gm_N^2$. 
In units where particle masses are constant, we have
\beq
 \dot{G}/G = -0.6 \pm 2.0\, (4.2) \times 10^{-12} \mbox{y}^{-1}, 
\eeq
where the averaging is performed over the last $12\times 10^{9}$y, and the bound should be reinterpreted for variations which are not linear in time. The absolute variation over this period is then bounded at $1\sigma$ as
\beq
 \Delta \ln G = (-0.7 \pm 2.4) \times 10^{-2}, \quad z=3.3.
\eeq

\subsection{Atomic clocks} \label{clocks}
Stringent bounds on the present time variation of the fine structure constant and the electron-proton mass ratio may be obtained by comparing different atomic transitions over periods of years in the laboratory \cite{Peik:2006xy}. A recent evaluation \cite{Blatt:2008su} of atomic clock data gives
\begin{align}
 d \ln \alpha / dt &= (-0.31 \pm 0.3)\times 10^{-15}\, \mbox{y}^{-1} \nonumber \\
 d \ln \mu / dt &= (1.5 \pm 1.7) \times 10^{-15}\, \mbox{y}^{-1}.
\end{align}

Fortier {\em et al.}\ \cite{Fortier:2007jf} obtain stronger bounds, $|\dot{\alpha}/\alpha| < 1.3\times 10^{-16}\,$y$^{-1}$, if other relevant parameters are assumed not to vary. If other atomic physics parameters are allowed to vary, this bound becomes considerably weaker, depending on a possible relative variation of the Cs magnetic moment and the Bohr magneton.
Direct comparison of optical frequencies may yield bounds at the level of $10^{-17}$ per year; limits on variation of $\alpha$ from this method are reported with uncertainty $2.3 \times 10^{-17}/$y \cite{Rosenband} but designated as preliminary. If these bounds are used then our limits from atomic clocks via $\alpha$ variation should be tightened by about an order of magnitude.

Extrapolating the results of \cite{Blatt:2008su} to the time of Oklo ($z=0.14$, $t = 1.8 \times 10^9\,$y) gives 
\begin{align}
  \Delta \ln \alpha &= (-0.56 \pm 0.54) \times 10^{-6}, \nonumber \\
  \Delta \ln \mu &= (-0.27 \pm 0.31) \times 10^{-5}.
\end{align}
The bound on $\alpha$ at this epoch is weaker than that from Oklo, Eq.~(\ref{Okloalpha}). The bound on $\mu$ cannot be directly compared, due to unquantified theoretical uncertainty in the Oklo bound. However, if we interpret Oklo as simply bounding $\Delta \alpha/\alpha$, we find that it provides the strongest bound on $\mu$ variation for all unified scenarios we consider (see Section \ref{sec:Unification}) except our scenario 3, where the high ratio $\Delta \ln \mu / \Delta \ln \alpha \simeq -325$ means that atomic clock bounds on $\mu$ are the most sensitive.
%--------------------

\section{Unification and relations between \\coupling variations}
\label{sec:Unification}

In this paper we consider the hypothesis that, for all redshifts, all fractional variations in the ``fundamental'' parameters $G_k$ (see section \ref{BBN_CMB}) are proportional to one nontrivial variation with fixed constants of proportionality. If the variation of the unified gauge coupling $\Delta \ln \alpha_X$ is nonvanishing, we may write
\beq \label{eq:dkdef}
	\Delta \ln G_k = d_k\Delta \ln\alpha_X
\eeq
for some constants $d_k$, assuming small variations. Different unification scenarios correspond to different sets of values for the ``unification coefficients'' $d_k$.
%\footnote{In the case where $\alpha_X$ does not vary, one can define $d'_k$ with respect to some other nonzero fundamental variation.} 
Considering the values of $\Delta\ln G_k$ as coordinates in an $N_k$-dimensional space, this assumption restricts variations to a single line passing through zero. The variation then constitutes exactly one degree of freedom. %, $\nu=1$. 
We will go beyond this hypothesis in a subsequent paper \cite{Part2} where we consider a specific model for which a fixed linear relation (\ref{eq:dkdef}) is not realized for all $z$.

\subsection{GUT relations}
\label{sec:GUTrels}
It is natural in this context to consider models with unification of gauge couplings (GUTs).
% where the cosmon might in general couple to four quantities. 
These have the property that variations of the Standard Model gauge couplings and mass ratios can be determined in terms of a smaller set of parameters describing the unified theory and its symmetry breaking. Hence, if nonzero variations in different observables are measured at similar redshifts, models of unification may be tested without referring to any specific hypothesis for the overall cosmological history of the variation. We need only assume that for a given range of $z$ the time variation is slow and approximately homogeneous in space, hence $\Delta \ln\alpha_X$ depends only on redshift $z$ to a good approximation.
The relevant unified parameters are the unification mass $M_X$ (relative to the Planck mass), the unified gauge coupling $\alpha_X$ defined at the scale $M_X$, the Higgs v.e.v.\ $\vev{\phi}$ and, for supersymmetric theories, the soft supersymmetry breaking masses $\tilde{m}$, which enter in the renormalization group (RG) equations for the running couplings. Then, for the variations at any given $z$ we can write 
%Using $l$ as multiplier factor according to Eq.~\eqref{eq:defln}, we define
\beq \label{eq:unifdefs}
\Delta \ln \frac{M_X}{M_{\rm P}} = d_M l, \quad \Delta \ln \alpha_X = d_X l, \quad \Delta \ln \frac{\vev{\phi}}{M_X} = d_H l, \quad \Delta \ln \frac{\tilde{m}}{M_X} = d_S l,
\eeq
where $l(z)$ is the ``evolution factor'' introduced for later convenience. If $\alpha_X$ varies nontrivially we may normalise $l$ via $d_X=1$. In supersymmetric theories we set $\alpha_X = 1/24$, in nonsupersymmetric theories we set $\alpha_X = 1/40$ and $d_S \equiv 0$ \cite{Dent03}. 

We make the simplifying assumption that the masses of Standard Model fermions all vary as the Higgs v.e.v., {\em i.e.}\ Yukawa couplings are constant at the unification scale: 
\beq \label{eq:fermionMX}
 \Delta \ln \frac{m_e}{M_X} = \Delta \ln \frac{\delta_q}{M_X} = \Delta \ln \frac{\hat{m}}{M_X} = \Delta \ln \frac{m_s}{M_X}
 = \Delta \ln \frac{\vev{\phi}}{M_X}.%, \qquad (Q^2=M_X^2).
\eeq
Here we neglect the effects induced by a variation of $\alpha_X$ on the RG running of fermion masses, and consider the quark masses defined at an appropriate RG scale for low-energy observables. We have explicitly calculated the effect of varying $\alpha_3(M_X)$ on the running of quark masses: for low-energy observables such as $m_q(Q^2)/\Lambda_c$ one should consider an RG scale $Q^2$ that is fixed relative to $\Lambda_c$. Thus the variation of $m_q(Q^2)/m_q(M_X^2)$ is entirely due to the dependence on $\alpha_3(M_X)$, which is suppressed by a loop factor $\alpha_X/\pi$ compared to the nonperturbative dependence of $\Lambda_c/M_X$ on $\alpha_X$.\footnote{We find $\Delta \ln (\bar{m}_q(Q^2)/\bar{m}_q(M_X^2)) = 2/7\Delta\ln \alpha_X \simeq (9\alpha_X/7\pi)\Delta \ln (\Lambda_c/M_X) $ under variation of $\alpha_X$, where $\bar{m}_q$ is the running quark mass and $Q^2=\,$const$\cdot\Lambda_c^2$.} Such effects enter at the order of 1\% correction, which is already smaller than our uncertainties in hadronic and nuclear physics.
%\textbf{: Support this by giving a explicit computation / orderof magnitude analysis}. 
With the assumption (\ref{eq:fermionMX}) one finds for the QCD scale \cite{Wetterich:2002ic,Dent03}
\beq
	\frac{\Delta \ln (\Lambda_c/M_X)}{l} = \frac{2\pi}{9\alpha_X}d_X + \frac{2}{9}d_H + \frac{4}{9}d_S
\eeq
and for the fine structure constant, %we obtain \cite{Dent03,Wetterich:2002ic}
\beq
	\frac{\Delta \ln \alpha} {l} = \frac{80 \alpha}{27 \alpha_X} d_X + \frac{43}{27}\frac{\alpha}{2\pi} d_H + \frac{257}{27} \frac{\alpha}{2\pi} d_S.
\eeq
Similar values with somewhat different assumptions were found earlier \cite{CalmetLangacker,DentF}.

For the nucleon mass we include possible strange quark contributions.\footnote{In our previous paper \cite{DSW07} we assumed $m_s/\Lambda_c =\, {\rm const}$, here we include the roughly known strange contribution to the proton mass. For BBN, the difference in the final dependence is less than $3\%$ and hence much lower than the model uncertainty ({\it e.g.}\ for nuclear binding energies).} The uncertainty in the strangeness content is an indicator of the overall uncertainty that may arise due to $m_s$ variation. We obtained \cite{DSW07}
\begin{align}
 	\Delta \ln \frac{m_N}{\Lambda_c} &= 0.048 
	\Delta \ln \frac{\hat{m}}{\Lambda_c} + (0.12 \pm 0.1) 
	\Delta \ln \frac{m_s}{\Lambda_c}, \label{eq:strangeness}\\
 	\Delta \ln \frac{\delta_N}{\Lambda_c} &= -0.59 \Delta \ln \alpha + 
	1.59 \Delta \ln \frac{\delta_q}{\Lambda_c},
\end{align}
where $\delta_N\equiv m_n-m_p$, and thus
\begin{align} 
	\frac{\Delta \ln \mu}{l} &= (0.58 \mp 0.08) \frac{d_X}{\alpha_X} + 
	(0.37 \mp 0.05) d_S + (-0.65 \pm 0.09) d_H, \\
	\frac{\Delta \ln (G m_N^2)}{l} &= 2 d_M + (1.16 \mp 0.17)
	\frac{d_X}{\alpha_X} + (0.74 \mp 0.11) d_S + (0.71 \pm 0.19) d_H, \label{DelGmNsquared}
\end{align}
where the upper or lower signs correspond to the positive or negative signs in Eq.~\eqref{eq:strangeness} respectively. 

The largest contribution to variations of the proton g-factor $g_p$ has been argued to arise from the pion loop \cite{Murphy:2003mi}, yielding at first order a dependence on the light quark mass of
\begin{align}
  \Delta \ln g_p &\simeq -0.087 \Delta \ln \hat{m}/\Lambda, \nonumber \\ 
  \frac{\Delta \ln g_p}{l} &\simeq 0.06 \frac{d_X}{\alpha_X} - 0.07 d_H + 0.04 d_S.
\end{align}
Hence the variations of observables including $g_p$ are
\begin{align}
 \frac{\Delta \ln x}{l} &= (-0.48 \pm 0.08) \frac{d_X}{\alpha_X} + (0.59 \mp 0.09) d_H + (-0.31 \pm 0.05) d_S \nonumber \\
 \frac{\Delta \ln y}{l} &= 0.10 \frac{d_X}{\alpha_X} -0.06 d_H + 0.06 d_S \nonumber \\
 \frac{\Delta \ln F}{l} &= (1.04 \mp 0.13) \frac{d_X}{\alpha_X} + (-1.08 \pm 0.14) d_H + (0.65 \mp 0.08) d_S \nonumber \\
 \frac{\Delta \ln F'}{l} &= (-0.54 \pm 0.08) \frac{d_X}{\alpha_X} + (0.65 \mp 0.09) d_H + (-0.35 \pm 0.05) d_S.
\end{align}
We have now expressed the variations accessible to observation in terms of three (four) variables: $l$, $d_X$, $d_H$ (and $d_S$), where one parameter may be eliminated by normalization. Different unified scenarios will be characterized by different relations among these parameters. 
%Our strategy of analysis is explained in more detail in Section~\ref{Strategy}.

Most data points are upper bounds on a possible variation, with the exception of two epochs. First, we consider specifically whether claimed nonzero variations of $\alpha$ \cite{Murphy:2003mi} and $\mu$ \cite{Reinhold:2006zn} at redshift $2$--$3$ are compatible with one another, since the ratio of their fractional variations is predicted in each scenario.

Second, we consider whether there is an indication of nonzero variation at BBN. For no variation at BBN we obtain $\chi^2 = 17.9$ for 3 measured abundances (\hefo, D, \lise). This discrepancy between theory and observation is exclusively due to \lise. (Considering only \hefo\ and D, the value of $\chi^2$ is $0.24$.) If we wish to solve or ameliorate the ``lithium problem'' by a nonzero variation, we will require $\chi^2 / \nu$ to be not much larger than unity, taking $\nu = 2$ as appropriate for one adjustable parameter. If there is no significant range where the three abundances have a $2\sigma$ fit ($\chi^2/\nu\leq 4$) then we give up the hypothesis that the \lise\ problem is solved by coupling variations and instead assume that the observed depletion %of \lise\ 
is due to some astrophysical effect. In this case we consider only D and \hefo\ abundances as observational bounds on the size of variations at BBN.

We will now investigate six different scenarios for the variation of the grand unified parameters $\alpha_X$, $M_X/M_{\rm P}$, $\vev{\phi}/M_X$ and $\tilde{m}/M_X$. These will fix the unification coefficients $d_k$. For each unified scenario we display the $z$-dependence of the fractional variation (Figs~1-7). Each figure shows the available information from observations of different couplings, interpreted as constraints on the variation of a single parameter. These figures are one of the main results of our paper.

\begin{figure}[p]
\begin{center}
\includegraphics[width=13.5cm]{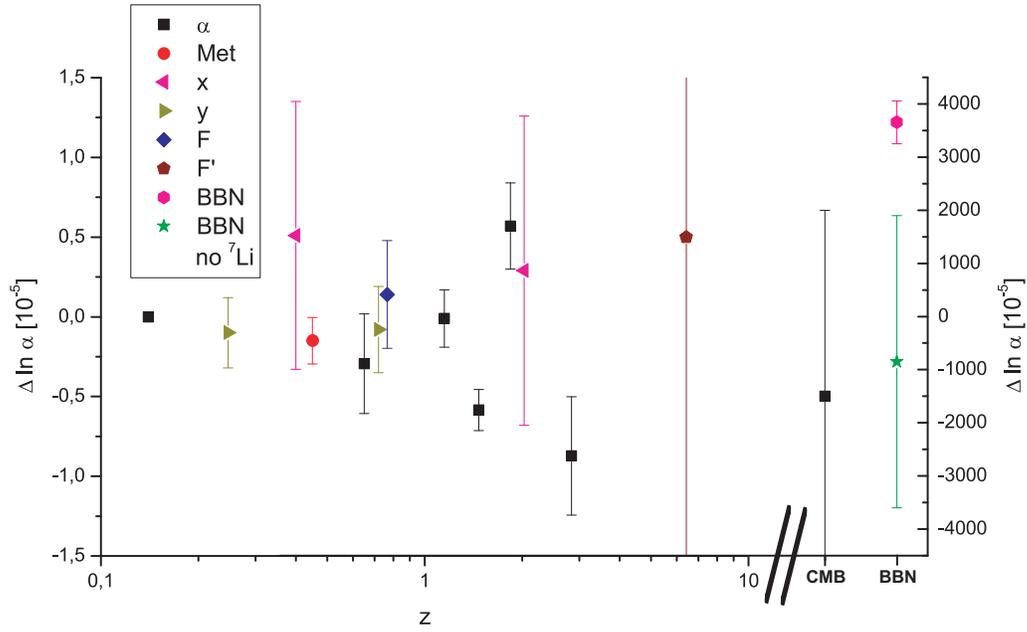}
\end{center}
\caption{Variations for varying $\alpha$ alone. Only observations constraining $\alpha$ variation are shown; the BBN fit including \lise\ is poor ($\chi^2/\nu\ge 7.8/2$) hence we also display a conservative bound from \hefo\ and D abundances neglecting \lise.}
\label{fig:alphaonly}
\end{figure}

\paragraph{Varying $\alpha$ alone}
%\subsection{Scenario ``0'': Varying $\alpha$ only}
Before describing the six different grand unified scenarios, we consider a variation of the fine structure constant $\alpha$ alone. Clearly here we are unable to account for any nonzero variation in $\mu$ or other quantities independent of $\alpha$. The cosmological history is dominated by the nonzero variation of the M$\alpha$ values at redshifts $z\simeq 1$ to $4$.
We find that there is almost no $2 \sigma$ match of the BBN values ($\chi^2 / \nu \ge 3.9$): the 2-sigma range is
\beq
	3.25 \% \ge \Delta \ln \alpha_{BBN} \ge 4.06 \%.
\eeq
Hence it seems unlikely that the ``lithium problem'' can be solved by a variation of $\alpha$ alone. If we regard the \lise\ discrepancy as due to systematic or astrophysical effects 
%in relating observations to the primordial abundance, 
we can set a conservative bound on $\alpha$ variation from \hefo\ and D abundances \cite{DSW07}
\beq
	-3.6\% \ge \Delta \ln \alpha_{BBN} \ge 1.9\%,
\eeq
where we imposed that neither the D nor \hefo\ abundance should deviate by more than $2\sigma$ from observational values. See Fig.~\ref{fig:alphaonly} for a summary of the bounds in this case.
% We cannot compare this result with the results of Nollett because he makes no combined analysis of all 3 abundances (as there is no 1 sigma fit). Bergstroem has no eta value and hence also does not give useful bounds.
%In this scenario we may derive an upper limit on the present-day rate of variation from the Oklo bound, assuming that the rate has not accelerated.
%
\begin{figure}[p] 
\begin{center}
\includegraphics[width=13.5cm]{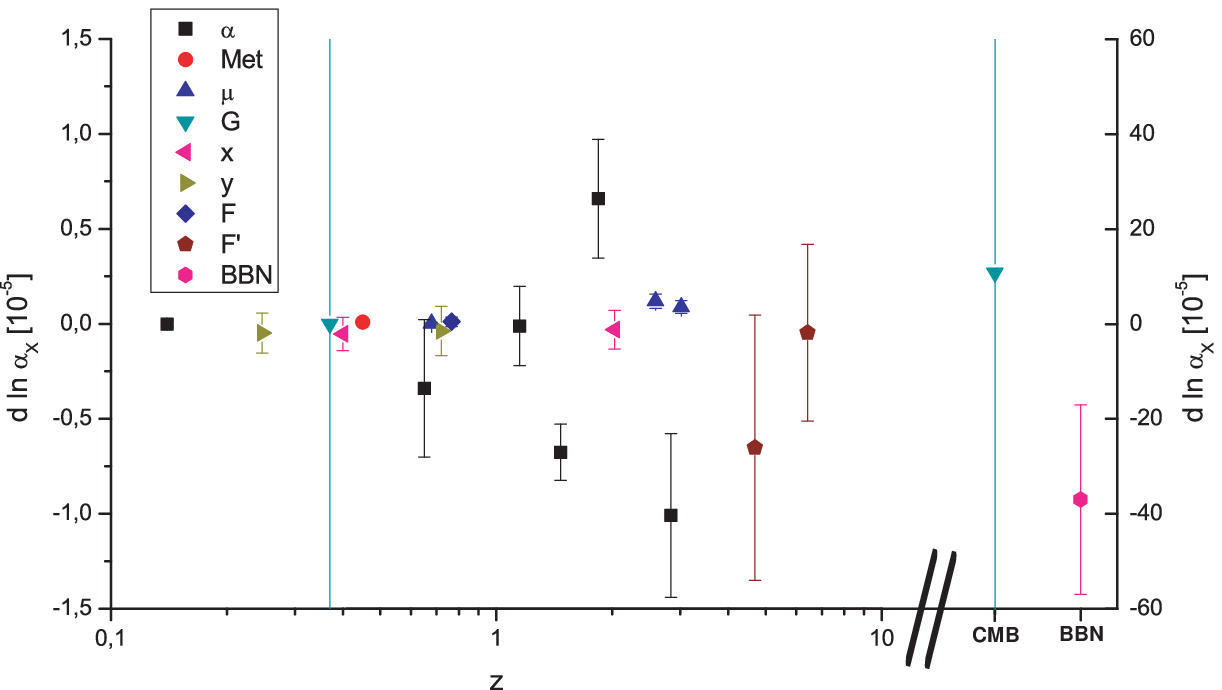}
\end{center}
\caption{Variations for scenario 2; BBN bounds are $2 \sigma$ bounds.}
\label{fig:scen2}
\end{figure} 

\subsection{Scenario 1: Varying gravitational coupling}
In this scenario we have only $d_M$ nonvanishing, 
\beq
	d_H = d_S = d_X = 0,
\eeq
therefore
\beq
	\Delta \ln \frac{M_X}{M_{\rm P}} = \frac{1}{2} \Delta \ln G \Lambda_c^2.
\eeq
We find that there is no value of $\Delta \ln G \Lambda_c^2$ for which BBN is consistent with the three observed abundances within $2 \sigma$. The best fit values are $\chi^2 / \nu \ge 7.7$ for no variation of $m_N/M_{\rm P}$ at CMB and  $\chi^2 / \nu \ge 5.9$ if the variation of $m_N/M_{\rm P}$ has the same size at BBN and CMB.
Assuming that the discrepancy in the \lise\ abundance is due to some other effect, we find the allowed region of variation of $G$ at BBN under which primordial D and \hefo\ abundance lie within the observed range at $1 \sigma$ ($2 \sigma$),
\beq
	-5\% \; (-13\%) \le \Delta \ln G \Lambda_c^2 \le 12\% \; (22\%)
\eeq
If the variation of $m_N/M_{\rm P}$ has the same size at BBN and CMB one finds
\beq
	-4\% \; (-11\%) \le \Delta \ln G \Lambda_c^2 \le 10\% \; (16\%).
\eeq

The bounds on time variation of $G\Lambda_c^2$ are much weaker than for many other varying couplings.
%Since the data are not consistent with a significant variation at any epoch, we do not consider this scenario further. 
This scenario also predicts a vanishing value of $\eta$ in E{\" o}tv{\" o}s experiments. Thus, to any one of the following scenarios we may add an additional nonzero $d_M$ of similar size to $d_X$, $d_H$ or $d_S$ without changing the results significantly.

\subsection{Scenario 2: Varying unified coupling}
In the first GUT scenario without SUSY we consider the case when only $d_X$ is nonvanishing,
\beq
 d_X=1,\qquad d_H = d_S = d_M = 0, \qquad \alpha_X = 1/40.
\eeq
Within a supersymmetric theory the same relations will apply except that $\alpha_X=1/24$ and the variations of observables are scaled by a factor $24/40$ relative to $\Delta \ln \alpha_X$: we designate this as Scenario 2S.
In both cases we find here
\beq
 \frac{\Delta \ln \mu}{\Delta \ln \alpha} \simeq 27. 
\eeq
It is then highly unlikely for the nonzero M$\alpha$ result for variation of $\alpha$ 
% \cite{Murphy:2003mi} 
to coexist with the determination of $\mu$ at redshift around $3$ \cite{Reinhold:2006zn}, even if the latter is interpreted as an upper bound on the absolute size of variation \cite{Wendt08}.

For the BBN fit, we find without SUSY (excluding modifications of the baryon fraction $\eta$ due to varying $m_N$) 
%(including \lise, $\nu=2$) 
no range of values fitting at $1\sigma$ level ($\chi^2 / \nu \ge 2.3$). At $2\sigma$ the abundances, including \lise, become consistent for the range
\beq
 -5.7 \times 10^{-4} \le \Delta \ln \alpha_X \le -1.7 \times 10^{-4} \qquad (2 \sigma).
\eeq
If one includes a variation of $m_N$ at the time of CMB with the same magnitude as at BBN the result remains unchanged ($\chi^2 / \nu \ge 2.45$), with the same $2 \sigma$ range. For this scenario we may consider a nonzero variation at BBN, but more recent probes must all be viewed as increasingly tight null bounds.
\begin{figure}[p]
\begin{center}
\includegraphics[width=13.5cm]{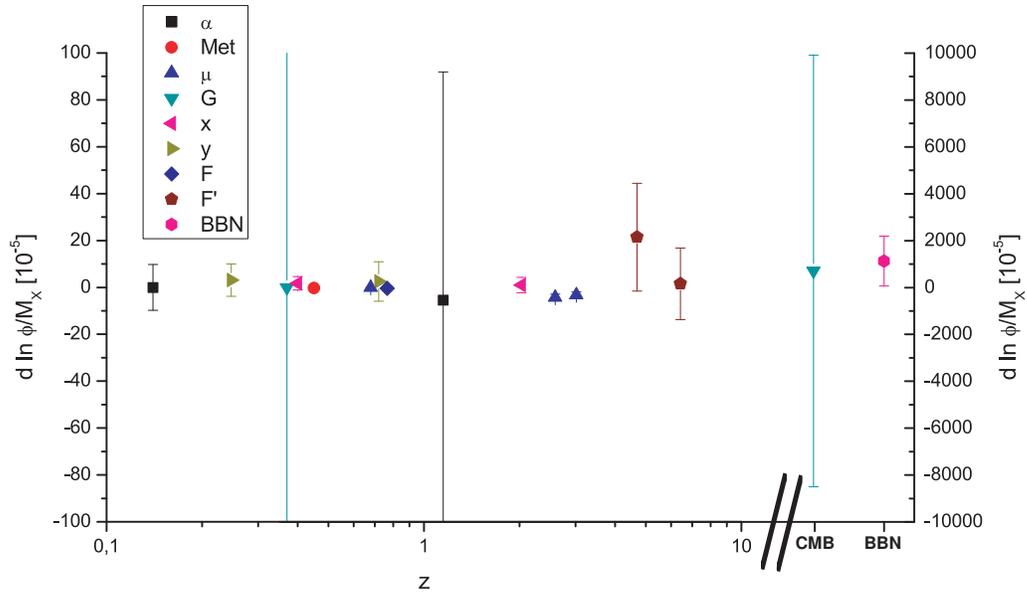}
\end{center}
\caption{Variations for scenario 3; BBN bounds are $2 \sigma$. Note that due to the very large ratio $\Delta \ln \mu/ \Delta \ln \alpha$ in this scenario, points indicating any nonzero variation of $\alpha$ fall well outside the range of the graph.}
\label{fig:scen3_small}
\end{figure} 

\subsection{Scenario 3: Varying Fermi scale}
In this scenario we consider the case when the variation arises solely from a change in the Higgs expectation value relative to the unified scale, thus only $d_H$ is nonzero:
\beq
 d_H=1,\quad d_S = d_M = d_X = 0, \qquad \alpha_X = 1/40.
\eeq
This scenario implies
\beq
 \frac{\Delta \ln \mu}{\Delta \ln \alpha} = -325.
\eeq
Whether we interpret the determination of $\mu$ \cite{Reinhold:2006zn} as a detection or an upper bound, any variation in $\alpha$ at large redshift case should be orders of magnitude smaller than current observational sensitivity.

We find for BBN including \lise\ ($\nu=2$) no $1\sigma$ range ($\chi^2 / \nu \ge 1.95$) but 
\beq
6 \times 10^{-3} \le \Delta \ln \vev{\phi}/M_X \le 22 \times 10^{-3} \qquad (2 \sigma).
\eeq
A variation of $m_N$ at the time of CMB with the same magnitude as at BBN does not change this result.% remains unchanged ($\chi^2 / \nu \ge 1.96$) with the same $2 \sigma$ allowed range.

\subsection{Scenario 4: Varying Fermi scale and SUSY-breaking scale}
\begin{figure}[p]
\begin{center}
\includegraphics[width=13.5cm]{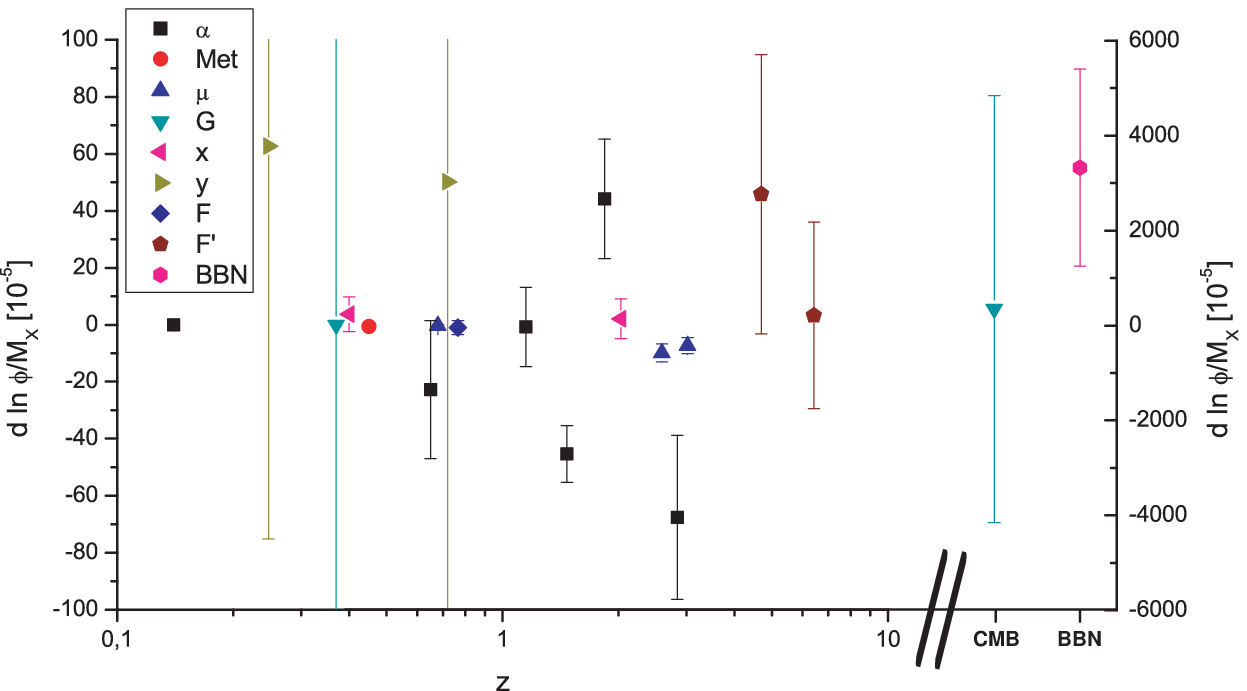}
\end{center}
\caption{Variations for scenario 4; BBN bounds are $2 \sigma$}\label{fig:scen4}
\end{figure}

This scenario corresponds to scenario 3, but includes supersymmetry and assumes that the mass-generating mechanism for SM particles and their superpartners gives rise to the same variation:
\beq
 d_M = d_X = 0, \quad d_S = d_H = 1, \qquad \alpha_X = 1/24.
\eeq
We find here
\beq
 \frac{\Delta \ln \mu}{\Delta \ln \alpha} = -21.5, 
\eeq
such that again the claimed nonzero variations in $\alpha$ and $\mu$ cannot be compatible and the variation in $\alpha$ at redshift $3$ must be below current sensitivities. We demonstrate this in Fig.~\ref{fig:scen4}, where we show for this scenario the bounds on the variable $d_Hl = \Delta \ln (\vev{\phi}/M_X)$ that arise from various observations. Since we have only one free variable we can plot all observations simultaneously as a function of redshift. Inspection ``by eye'' permits to judge if a smooth and monotonic evolution of $d_Hl$ is consistent or not.

We find for BBN including \lise ($\nu=2$) no $1\sigma$ fit ($\chi^2 / \nu \ge 1.60$), while at $2\sigma$ 
\beq
1.25 \times 10^{-2} \le \Delta \ln \vev{\phi}/M_X \le 5.4 \times 10^{-2} \qquad (2 \sigma).
\eeq
If one includes a variation of $m_N$ at the time of CMB with the same magnitude as at BBN the allowed range becomes slightly restricted ($\chi^2 / \nu \ge 1.72$),
\beq
1.20 \times 10^{-2} \le \Delta \ln \vev{\phi}/M_X \le 4.9 \times 10^{-2} \qquad (2 \sigma).
\eeq

\subsection{Scenario 5: Varying unified coupling and Fermi scale}
In this scenario we study a combined variation of the unified coupling and the Higgs expectation value:
\beq
 d_M = d_S = 0,\quad d_X=1, \quad d_H = \tilde{\gamma} d_X, \qquad \alpha_X = 1/40.
\eeq
The parameter $\tilde{\gamma}$ can be related to the parameter $\gamma \equiv \frac{ \Delta \ln \vev{\phi} / M_X}{\Delta \ln \Lambda_c / M_X}$ which was introduced in \cite{DSW07} via
\beq
	\gamma = \tilde{\gamma} \left( \frac{2\pi}{9 \alpha_X} + 
	\frac{2}{9} \tilde{\gamma} \right)^{-1}.
\eeq
%\beq
%  \tilde{\gamma} = \frac{2\pi}{9 \alpha_X} \left( \gamma^{-1} - \frac{2}{9} \right)^{-1} 
%\eeq
%
In \cite{DSW07} we examined the cases $\gamma = (0, 1, 1.5)$ which correspond to $\tilde{\gamma} = (0, 36, 63)$. Here we find that the best BBN fit is reached for $\tilde{\gamma} \approx 50$ with $\chi^2/\nu = 1.45$. Note that we have the freedom to adjust $\tilde{\gamma}$ such that nonzero variations of $\alpha$ and $\mu$ at redshift $\simeq 3$ are consistent with each other. 
We have
\beq
	\frac{\Delta \ln \mu}{\Delta \ln \alpha} = \frac{23.2 -0.65\tilde{\gamma}}
	{0.865 + 0.002\tilde{\gamma}}\,.
\eeq
We choose for illustration $\tilde{\gamma} = 42$, for which
\beq
  \Delta \ln \mu = -5.6\, \Delta \ln \alpha %\qquad (\tilde{\gamma} = 42)
\eeq
and the $2 \sigma$ contour for BBN is 
\beq
  7.5 \times 10^{-4} \le \Delta \ln \alpha_X \le 28 \times 10^{-4}.
\eeq
For a variation of $m_N$ at the time of CMB with the same magnitude as at BBN the fit becomes worse ($\chi^2 / \nu \ge 1.68$). However, a $2 \sigma$ fit to BBN is obtained over a wide range of $0 \le \tilde{\gamma} \le 26 $ (negative $\Delta \ln \alpha_X$) and $40 \le \tilde{\gamma} < \infty$ (positive $\Delta \ln \alpha_X$). 

\begin{figure}[p]
\begin{center}
 \includegraphics[width=13.5cm]{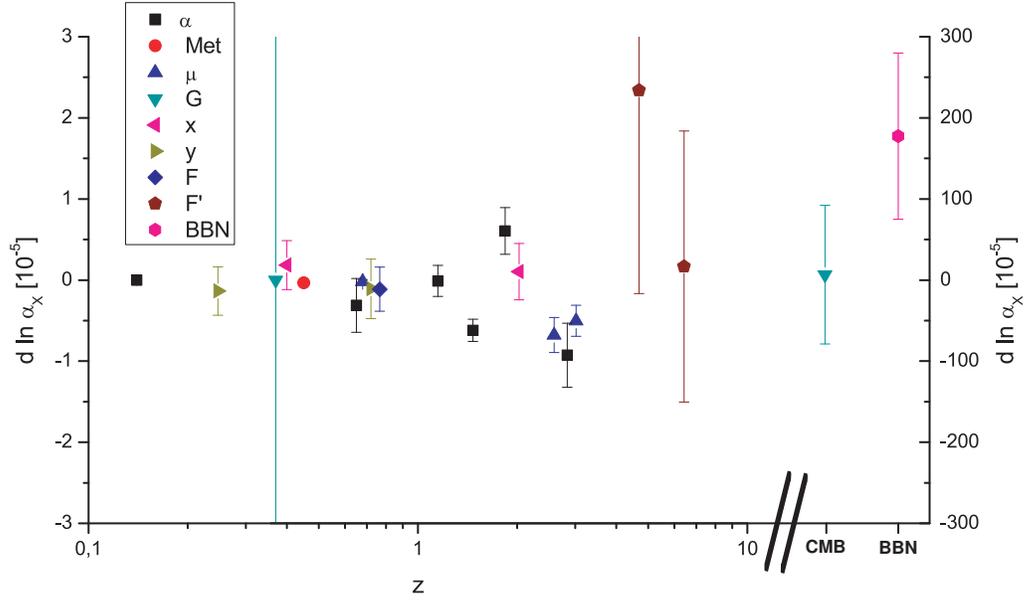}
%	\\
% \includegraphics[width=12cm]{scenario4_42_small.eps}
\end{center}
\caption{Variations for scenario 5, $\tilde{\gamma} = 42$; BBN bounds are $2 \sigma$} \label{fig:scen542}
\end{figure}

Assuming that the apparent \lise\ mismatch at BBN is due to systematic astrophysical effects, we may bound $\alpha_X$ with only D and \hefo\ abundances. Here we find at $1 \sigma$ 
\beq
  -5.5 \times 10^{-4} \le \Delta \ln \alpha_X \le 1.44 \times 10^{-3} 
\eeq
In Fig.~\ref{fig:scen542} we again plot simultaneously all observations for this scenario. This shows that the bound from BBN including \lise\ is not consistent with the claimed nonzero variations of $\alpha$ and $\mu$ for a monotonic evolution over $z$. 
%For further considerations, we will release the \lise\ bound and continue with the latter bound.
%
\begin{figure}[p]
\begin{center}
 \includegraphics[width=13.5cm]{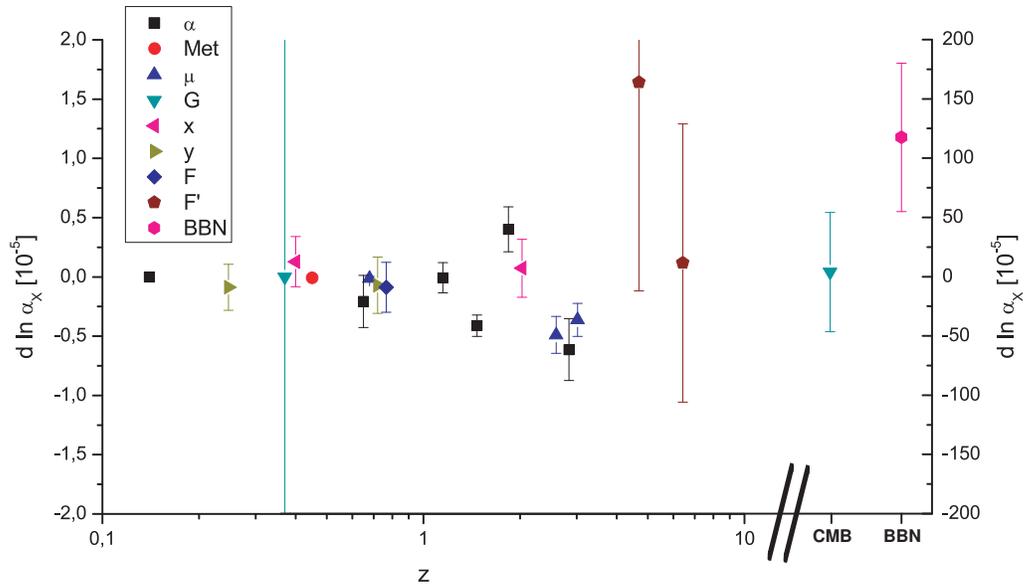} 
\end{center}
\caption{Variations for scenario 6, $\tilde{\gamma} = 70$; BBN bounds are $2 \sigma$} \label{fig:scen670}
\end{figure} 

\subsection{Scenario 6: Varying unified coupling and Fermi scale with SUSY}
In this scenario we study a combined variation of the unified coupling and the Higgs v.e.v.\ including SUSY, where as in Scenario 4 we tie the variations of the superpartner masses and Fermi scale together:
\beq
 d_M = 0,\quad d_X=1, \quad d_S \simeq d_H = \tilde{\gamma} d_X, \qquad \alpha_X = 1/24.
\eeq
Now the relation to $\gamma$ is modified as
\beq
  \gamma = \tilde{\gamma} \left( \frac{2\pi}{9 \alpha_X} + \frac{2}{3} \tilde{\gamma} \right)^{-1}
\eeq
%\beq
%  \tilde{\gamma} = \frac{2\pi}{9 \alpha_X} \left( \gamma^{-1} - \frac{2}{3} \right)^{-1} 
%\eeq
%
One may again adjust $\tilde{\gamma}$ to make nonzero variations in $\alpha$ and $\mu$ self-consistent. With
\beq
	\frac{\Delta \ln \mu}{\Delta \ln \alpha} = \frac{14 - 0.28\tilde{\gamma}}
	{0.52 +  0.013\tilde{\gamma}} \,,
\eeq
we find that a good fit to BBN is obtained over a large range of $\tilde{\gamma}$, ranging from $\tilde{\gamma} = 100$ to infinity with minimal $\chi^2 / \nu = 1.45$. This shows that the main effect in the SUSY model comes from the variation of the Higgs v.e.v. Including a variation of $m_N$ at the time of CMB with the same magnitude as at BBN the fits gets worse ($\chi^2 / \nu \ge 1.8$). A $2 \sigma$ fit can be obtained for $0 \le \tilde{\gamma} \le 28$ (for negative $\Delta \ln \alpha_X$ at BBN) and for $58 \le \tilde{\gamma} < \infty$ (positive $\Delta \ln \alpha_X$). 

\begin{figure}[htb]
\begin{center}
 \includegraphics[width=13.5cm]{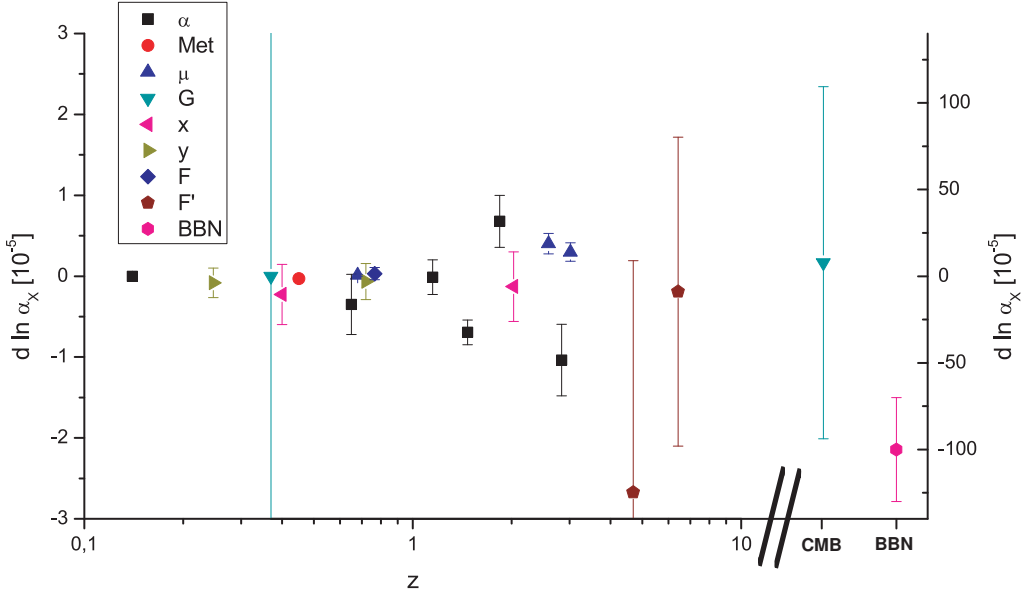}
\end{center}
\caption{Variations for scenario 6, $\tilde{\gamma} = 25$; BBN bounds are $2 \sigma$} \label{fig:scen625}
\end{figure}

First, we study the case $\tilde{\gamma} = 70$ for which 
\beq
  \Delta \ln \mu = -3.9 \Delta \ln \alpha \qquad (\tilde{\gamma} = 70)
\eeq
and BBN is fit with a $2 \sigma$ range
\beq
  5.5 \times 10^{-4} \le \Delta \ln \alpha_X \le 18 \times 10^{-4}.
\eeq
Neglecting \lise, we obtain a $1 \sigma$ bound from BBN
\beq
  -3.5 \times 10^{-4} \le \Delta \ln \alpha_X \le 9.3 \times 10^{-4}.
\eeq
%In Fig.~\ref{fig:scen670} we display the bound of BBN including \lise.
%
Secondly, we study the case $\tilde{\gamma} = 25$ where 
\beq
  \Delta \ln \mu = 8.3 \Delta \ln \alpha \qquad (\tilde{\gamma} = 25),
\eeq
and where the $2 \sigma$ contour for BBN is
\beq
  -13 \times 10^{-4} \le \Delta \ln \alpha_X \le -7 \times 10^{-4}.
\eeq
In this second case the Murphy $\alpha$ measurement and BBN point into the same direction. The difference between the two values of $\tilde{\gamma}$ can be seen from a comparison of Figs.~\ref{fig:scen670} and \ref{fig:scen625}.

\section{Epochs and evolution factors}
\label{sec:Epochs_and_evolutionFactors}

\subsection{Epochs}
In this section we group the information contained in Tables 1-3 and figures 1-7 into different cosmological epochs. This produces a first quantitative estimate of the possible time evolution for the various unified scenarios. The choice of epochs is somewhat arbitrary. Two epochs are singled out by events in early cosmology, namely the last scattering surface of CMB, and BBN. The very recent epoch comprises present day laboratory experiments and the Oklo natural reactor, for which a linear interpolation to the present rate of varying couplings seems reasonable. We further divide the observations at intermediate redshift into three epochs.
\begin{itemize}
\item \textbf{Epoch 1:} Today until Oklo\\
 Contains Oklo and laboratory measurements. For the laboratory measurements, we extrapolate the rate of change of the couplings to finite changes at the redshift $z=0.14$ ($t = 1.8 \times 10^9\,$y) of the Oklo event.
\item \textbf{Epoch 2:} $0.2 \le z \le 0.8$\\
 Contains absorption spectra and isotopic abundance measurements in meteorites. We chose a boundary $z=0.8$ since the Murphy dataset \cite{Murphy:2003mi} has relatively few systems around this redshift, %$z=0.8$, 
 making a natural division.
\item \textbf{Epoch 3:} $0.8 \le z \le 2.4$\\
 Contains several absorption spectra measurements. The end of the Tzana\-varis dataset \cite{Tzanavaris:2006uf} sets the cut at $z=2.4$.
\item \textbf{Epoch 4:} $2.4 \le z \le 10$\\
 Contains absorption spectra measurements and bounds on $G$ from neutron stars.
\item \textbf{Epoch 5:} CMB, $z \approx 1100$
\item \textbf{Epoch 6:} BBN, $z \approx 10^{10}$
\end{itemize}

\subsection{Evolution factors}
We define ``evolution factors'' $l_n$ for epochs $n=1,\ldots,6$ by
\beq \label{eq:defln1}
	\Delta \ln G_{k,n} = d_{k} l_n .
\eeq
For each unification scenario we will proceed to a quantitative estimate of $l_n$, shown in Table~\ref{tab:multiplierFactors}. The usefulness of considering the evolution factors $l_n$ is that the unknown (and possibly not monotonic) behaviour of the mechanism driving the coupling variations is rolled into a finite number of parameters. For a monotonic behaviour they satisfy $l_n<l_p$ whenever $z_n<z_p$. The basic assumption remains the proportionality $\Delta \ln G_k(z_n) = d_k l(z_n) = d_k l_n$, with constant unification coefficients $d_k$ independent of the epoch.
The normalization of $l_n$ is arbitrary, and we take for scenarios 2, 5 and 6
\beq
\label{eq:defEvolutionCoefficient}
l_n = \Delta \ln \alpha_{X,n},
\eeq
while for scenarios 3 and 4 we take
\beq
l_n = \Delta \ln (\vev{\phi}/M_X)_{,n}.
\eeq
For each epoch and scenario, we compute the evolution coefficients $l_n$ as a weighted average over the measurements in the epoch. The representative redshift $z_n$ is the average over the redshifts of observations inside the corresponding epoch. It is shown together with the resulting values for $l_n$ in table \ref{tab:multiplierFactors}. This table summarizes our results under the assumption of proportionality.

\paragraph{Rates of time variation in the present epoch} 
\ \\
For Epoch 1 we incorporate the laboratory measurements for rates of varying couplings by linear extrapolation in time to the Oklo redshift $z_1=0.14$. The logarithmic time derivatives may be approximated by linear interpolation
\beq
	\frac{\dot{G}_k}{G_k} = \partial_t \ln G_k \simeq - \frac{d_k l_1}{t_0-t_1},
\eeq
where $t_1 = 1.8 \times 10^9$y is the time %(counting the Big Bang as $t=0$) 
corresponding to the redshift $z_1 = 0.14$. 

\paragraph{Method of averaging}
\ \\
We evaluate the weighted average using all values listed in table \ref{tab:CollectionOfMeasurements}. This procedure may be quite problematic, since sometimes different observations are in manifest contradiction. We take the attitude that, given the possible presence of systematic effects both in spectroscopic determinations of nonzero coupling variations and in the primordial \lise\ abundance, a viable model need not fit all data points. However, even if any given nonzero claimed variation is actually due to systematic error, we still expect the size of the error to be comparable to the size of the claimed variation. Thus, such claims are most conservatively interpreted as bounds on the absolute magnitude of variation. The surviving nonzero variation(s), in addition to the null bounds at other epochs, define a set of evolution factors which must be satisfied by any explicit model of evolution. 

For some scenarios we therefore also evaluate the evolution factors that are obtained by considering that some of the claimed observations of nonzero variation may instead be due to an underestimated systematic error. These alternative evolution factors are given in square brackets, corresponding to the following replacements:\\
Scenario 5, $\tilde{\gamma} = 42$: Neglecting \lise-abundance at BBN\\
Scenario 6, $\tilde{\gamma} = 70$: Neglecting \lise-abundance at BBN\\
Scenario 6, $\tilde{\gamma} = 25$: Replacing the $\mu$ measurements of \cite{Reinhold:2006zn} by the conservative upper bound of \cite{Wendt08}.\\
In the case where $\alpha$ alone varies, since the fit including \lise\ is poor we calculate a $2\sigma$ range using observational central values and errors of D and \hefo\ abundances given in \cite{DSW07}.
\begin{table}
%\center
\hspace*{-1.5cm}
\begin{tabular}{|l|cccccc|}
\hline
\ \ Epoch        & 1    & 2     & 3    & 4    & 5      & 6  \\
\ \ \ \ \ $z_n$  & 0.14 &  0.53 &  1.6 &  3.8 & $10^3$ & $10^{10}$ \\
\hline
Scenario & $l_1\times 10^6$ & $l_2\times 10^6$ & $l_3\times 10^5$ & $l_4\times10^5$ &
	$l_5\times 10^4$ & $l_6\times 10^3$ \\
\hline
$\alpha$ only 
  & $-0.01 \pm 0.06$ & $-1.1 \pm 1.0 $  & $-0.26 \pm 0.10$ & $-0.85 \pm 0.37$&	
	$-150 \pm 350$ & $5 \pm 34$       \\
2 & $-0.1 \pm 0.1$   & $0.04 \pm 0.03$  & $-0.15 \pm 0.08$ & $0.10 \pm 0.03$ & 
	$0.9 \pm 14$   & $-0.37 \pm 0.20$ \\
3 & $ 4.1  \pm 4.8$  & $-1.5 \pm 1.2$   & $0.42 \pm 3.3$  & $-3.6 \pm 0.9$   & 
	$69 \pm 920$   & $14 \pm 8$       \\
4 & $ 3.9   \pm 8.5$ & $-3.4 \pm 2.7$   & $-8.4 \pm 5.1$  & $-8.7 \pm 2.1$   & 
	$31 \pm 450$   & $33 \pm 21$      \\
5, & $-0.02 \pm 0.18$ & $-0.24 \pm 0.18$ & $-0.25 \pm 0.10$ & $-0.61 \pm 0.13$ & 
	$0.6 \pm 8.6$  & $1.7 \pm 1.1$    \\
($\tilde{\gamma} = 42$) &&&&&& [$0.4 \pm 1.0$] \\
6, & $-0.02 \pm 0.12$ & $-0.10 \pm 0.07$ & $-0.17 \pm 0.07$ & $-0.44 \pm 0.10$ & 
	$0.3 \pm 5.0$  & $1.2 \pm 0.6$    \\
($\tilde{\gamma} = 70$) &&&&&& [$0.3 \pm 0.6$] \\
6, & $-0.12 \pm 0.18$ & $0.04 \pm 0.12$  & $-0.30 \pm 0.11$ & $0.29 \pm 0.08$  & 
	$0.7 \pm 10$   & $-1 \pm 0.3$     \\
($\tilde{\gamma}= 25$) &&&&[$-0.43 \pm 0.28$]&&\\
\hline
\end{tabular}
\caption{Redshifts and evolution factors for each epoch, for the scenarios defined in section \ref{sec:Unification}. In the first row the values of $l_n$ give the fractional variation of $\alpha$; in Scenarios 2, 5 and 6 that of $\alpha_X$; and in 3 and 4 that of $\vev{\phi}/M_X$. Values in brackets give, for BBN ($l_6$) the evolution factors neglecting \lise; or for $l_4$, the evolution factor with the $\Delta \mu/\mu$ value of \cite{Reinhold:2006zn} substituted by that of \cite{Wendt08}.}
\label{tab:multiplierFactors}
\end{table}

\subsection{Monotonic evolution with unification}
\label{sec:monotony}
Here we briefly summarize whether the unified scenarios we consider can be consistent with a monotonic evolution of the single underlying varying parameter, based on the evolution factors $l_i$ found in Table~\ref{tab:multiplierFactors}.

\paragraph{Varying $\alpha$ only}
\ \\
Although variation of $\alpha$ alone does not help to account for deviation of BBN abundances from standard theory, or for any nonzero variation of $\mu$, the cosmic history is interesting due to the significant nonzero value in Epochs 3 and 4. The Oklo bound in Epoch 1 restricts the present time variation to $3.7\times 10^{-17}\,{\rm y}^{-1}$ (assuming no acceleration of $\partial_t \alpha$). 

\paragraph{Scenario 2}
\ \\
Scenario 2 favours a negative variation of $\alpha_X$ at BBN, and a negative variation may also fit the M$\alpha$ results. However, the Reinhold $\mu$ measurement indicates a positive, but much smaller, variation. We keep the R$\mu$ results, which dominate the weighted average due to their small error on $\Delta \ln \alpha_X$, to obtain $l_4$. The ratio $\Delta \ln \mu / \Delta \ln \alpha = 27$ makes this scenario unlikely to fit the reported signal of nonzero $\Delta \alpha$.

\paragraph{Scenario 3}
\ \\
In scenario 3 a positive variation of $\vev{\phi}/M_X$ is favoured by BBN. The high ratio \linebreak $\Delta \ln \mu / \Delta \ln \alpha \simeq -325$ makes the bounds obtained on a variation of $\mu$ strongly inconsistent with the claimed size of variation of $\alpha$. %Hence this scenario also appears unlikely. 
We keep the Reinhold {\it et al.}\ values to obtain $l_4$, which again dominate the results.

\paragraph{Scenario 4}
\ \\
In this scenario, the ratio $\Delta \ln \mu / \Delta \ln \alpha = -22$ is again large and makes any observation of significant nonzero $\Delta \ln \alpha$ unlikely. Both the M$\alpha$ and the R$\mu$ measurements point in opposite direction to BBN; however the two spectroscopic observations are also inconsistent with each other, within the scenario. Again, we keep the R$\mu$ results which dominate the determination of $l_4$ 
%einhold values and the results obtained for  are dominated by the  measurements 
due to the small error.

\paragraph{Scenario 5, $\tilde{\gamma} = 42$}
\ \\
In this scenario the variation of $\alpha_X$ favoured by BBN is positive ($l_6 = (1.7 \pm 1) \times 10^3$), however both nonzero variations from spectroscopic data M$\alpha$ and R$\mu$ require negative variations. With $\Delta \ln \mu / \Delta \ln \alpha = -6$ the spectroscopic measurements appear consistent with each other. Hence one would require some non-monotonic evolution to fit nonzero variations both at BBN and at moderate $z$.
%, which show a change in sign of variation between $z \approx 4$ and BBN. 
Thus in Table~\ref{tab:multiplierFactors} we have also evaluated $l_6$ using only the constraints given by D and \hefo\ (in brackets).

\paragraph{Scenario 6, $\tilde{\gamma} = 70$}
\ \\
As in the preceding scenario, BBN favours a positive variation in $\alpha_X$, but M$\alpha$ and R$\mu$ favour negative. Again, Fig.~\ref{fig:scen670} may suggest a non-monotonic evolution. 
%In this context we note that the points representing the low-, mid- and high-$z$ parts of the M$\alpha$ dataset represent averages over considerable ranges of redshift, for which a model with oscillations on a time scale shorter than the age of the Universe would tend to give zero or inconsistent results. 
Fitting to BBN including \lise\ we would obtain $l_6 = (1.2 \pm 0.6) \times 10^{-3}$; table~\ref{tab:multiplierFactors} also displays in brackets the value of $l_6$ obtained from D and \hefo\ bounds only.

\paragraph{Scenario 6, $\tilde{\gamma} = 25$}
\ \\
In this scenario, both BBN and the M$\alpha$ signal favour a negative variation of $\alpha_X$, whereas the R$\mu$ observations point towards a positive variation. Following the argument of Wendt {\it et al}.~\cite{Wendt08}, we substitute the R$\mu$ value by the null constraint $|\Delta \mu / \mu| \le 2.5 \times 10^{-5}$ \cite{Wendt08} to obtain the bracketed value of $l_4$ in Table~\ref{tab:multiplierFactors}. In this scenario the evolution factors show a crossover from negligible variation at low redshift, to strong and monotonically increasing negative variation at $z \approx 2$. 
%Such a behaviour is favoured by crossover quintessence models \cite{WetterichCrossoverQ}.

%\section{Further remarks and summary}

\subsection{Tension between the \lise\ problem and variation of $\mu$}
\label{sec:TensionLithiumMu}
Measurements of the primordial \lise\ abundance show that the BBN abundance needs to decrease below the standard value to fit the observations, whereas the Reinhold $\mu$ measurement indicates $\mu$ to increase at $z\simeq 3$. We find that for all our unification scenarios the sign of the dependence on the fundamental parameter is the same for $\mu$ and \lise. Moreover, the coefficients of this dependence are nearly identical up to a common factor; hence the induced variations for $\mu$ and \lise\ point in the same direction, in contradiction to the tendency inferred from the observations. For example, for scenario 5 we find
\begin{align}
 \Delta \ln \mu &= (23.2 - 0.65 \tilde{\gamma}) \Delta \ln \alpha_X, \nonumber \\ 
 \Delta \ln \mbox{\lise} &= (1692 - 49 \tilde{\gamma}) \Delta \ln \alpha_X.
\end{align}
These expressions change sign at $\tilde{\gamma} = 35.7$ and $34.5$, respectively. For a monotonic evolution, there is no possibility to have both a significant variation of $\mu$ and a variation of opposite sign in the \lise\ abundance. (In the regime $\tilde{\gamma} \approx 35$ there is no $2 \sigma$ fit to BBN.) A similar result can be found for scenario 6 (including the SUSY partner mass dependence, which shows the same sort of degeneracy). Note that scenario 2 and 3 are just limiting cases of scenarios 5 and 6.

The main reason for this behaviour is that variations of \lise\ and $\mu$ are dominated by the variations of $\hat{m}/\Lambda_c$ and $m_e/\Lambda_c$, respectively, with the same sign of prefactor. This degeneracy can be broken if $m_e$ varies differently from the quark masses, a possibility that we do not consider in this paper. For our scenarios with constant $\hat{m}/m_e$, the conflict between a monotonic time evolution and the $\mu$- and \lise-observations is reflected in the opposite signs of $l_4$ and $l_6$.

\begin{figure}[tb]
\begin{center}
 \includegraphics[width=13.5cm]{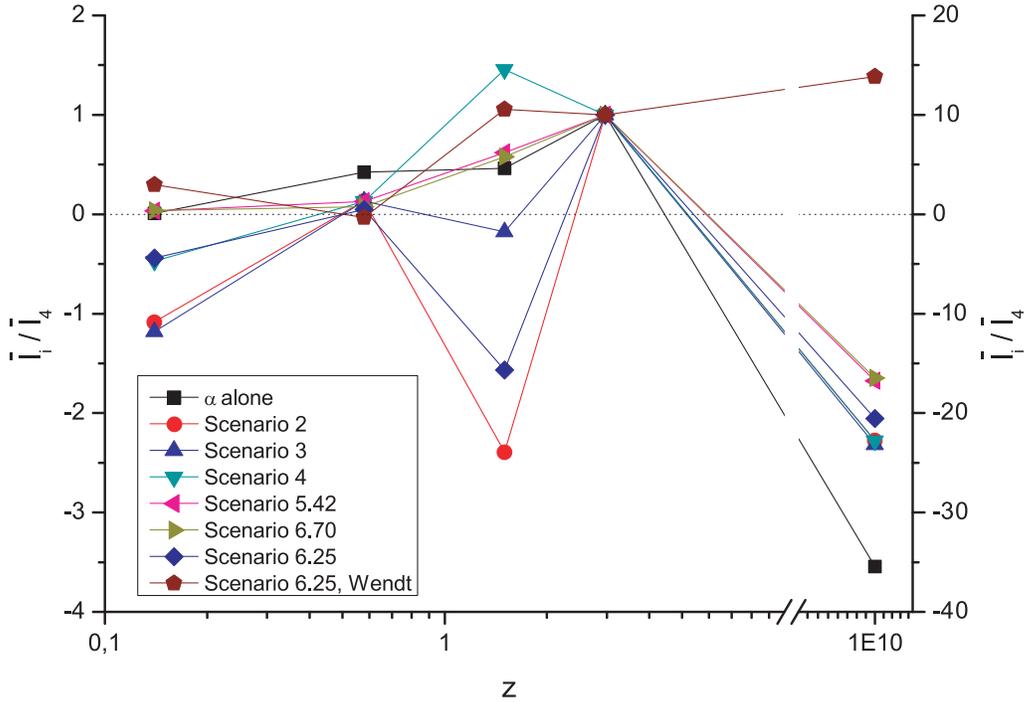}
\end{center}
\caption{Normalised evolution factors $\bar{l}_i/\bar{l}_4$ for each scenario, where $\bar{l}_i\equiv l_i/\ln(1+z_i)$.}
\label{fig:lionl4}
\end{figure}

This observational tension for monotonic behaviour is clearly depicted in Fig.~\ref{fig:lionl4}, 
where we plot simultaneously the averaged observational values of evolution factors $l_i/\ln(1+z_i)$, normalized to $l_4/\ln(1+z_4)$. For Scenario 6, $\tilde{\gamma}=25$, we also display the result obtained by substituting the Wendt {\it et al.}\ value of $\mu$ variation for that of \cite{Reinhold:2006zn}.
The factor $\ln(1+z_i)$ is introduced as a convenient normalization to avoid compressing the scale of variations excessively in recent epochs.\footnote{In quintessence-like theories, if the scalar field contributes a constant fraction of the total energy density of the Universe, as in so-called ``tracker'' models, the evolution of the field is typically also proportional to $\ln(1+z)$. This is an additional motivation for our normalization.} For the purpose of a quick inspection we have omitted the error bars, which are of course necessary for a quantitative interpretation.

\subsection{Special values of $\tilde{\gamma}$}
\label{sec:Specials}
In Scenarios 5 and 6 there is a value of $\tilde{\gamma}$ for which $\Delta \ln \vev{\phi}/\Lambda_c$ vanishes. For these values, Standard Model physics undergoes an overall multiplicative shift of energy scale under variation of $\alpha_X$, up to variations of perturbative, dimensionless couplings: specifically the Yukawa couplings (whose variation we have generally neglected) and $\alpha$. The significant observable effects arising from variation of SM fermion masses relative to $\Lambda_c$, which dominate in most unified scenarios, are largely absent, and the low-energy phenomenology is very similar to the case of varying $\alpha$ only. In particular the \lise\ problem at BBN is not addressed and the variation of $\mu$ is smaller than that of $\alpha$. 

The required values are $\tilde{\gamma}=2\pi/7\alpha_X \simeq 36$ in the case without SUSY ($\alpha_X\simeq 1/40$); or $\tilde{\gamma}= 2\pi/3\alpha_X\simeq 50$ with SUSY ($\alpha_X\simeq 1/24$) when the superpartner masses vary with the Fermi scale, $d_S=d_H=\tilde{\gamma}d_X$. From a low-energy point of view these values appear as fine-tuning, however it is conceivable that they would arise from some specific mechanisms of electroweak symmetry-breaking or SUSY-breaking.

\section{Summary / Conclusions}
\label{sec:Conclude}
Within Grand Unified Theories, different measurements of the variation of fundamental constants can be consistently reduced to a variation of a 
%single fundamental parameter. 
few ``unification parameters'', namely the unification scale $M_X/M_{\rm P}$, gauge coupling $\alpha_X$, the Fermi scale $\vev{\phi}/M_X$ and SUSY-breaking masses $\tilde{m}/M_X$. We define various GUT-scenarios for varying couplings by the assumption of proportionality of fractional variations of the unification parameters.

Assuming that couplings really vary, this is a way of excluding such GUT scenarios by demanding consistency of the implied variations. The assumption of
proportionality permits us to project all observations into constraints on a common evolution factor $l(z)$ for each scenario. We show that different GUT scenarios yield different time evolutions of $l(z)$ assuming that certain claimed measurements of varying constants are correct. We confirm that ``simple'' models which have only one fundamental parameter varying ($\alpha_X$ or $\vev{\phi}/M_X$) result in inconsistent variations. However, combined variations of these two parameters, as described in scenarios 5 and 6, lead to results more consistent with the possible quintessence-induced time variations of fundamental couplings which we investigate in \cite{Part2}. %However, in every case we had to drop one claimed non-zero variation, which shows that there is tension between the different claims. 

%These considerations show that variations of constants, if they occurred, provide a powerful tool for testing extensions of the Standard model or, vice versa, allow us to control consistency of claimed variations under the demand of unification.

Specifically, one may ask whether the claimed observations of variations in $\alpha$ \cite{Murphy:2003mi} and $\mu$ \cite{Reinhold:2006zn} are mutually consistent, and whether they are consistent with an explanation of the apparent primordial \lise-depletion by varying couplings. Within a hypothesis of constant Yukawa couplings, which results in identical fractional variations of all quark and lepton masses, we investigated arbitrary variations of $\alpha_X$, $\vev{\phi}/M_X$ and $M_X/M_{\rm P}$. For scenarios with supersymmetry we also assumed that the SUSY-breaking masses vary proportional to $\vev{\phi}$, but the effect of such a variation only appears at higher order and is probably not crucial. 

We have not found a scenario with a monotonic time evolution $l(z)$ that makes all three signals or hints of variation mutually consistent. A monotonic evolution requires either to discount one of the ``signals'' by substantially increasing its uncertainty, or to alter our assumptions by including additional time variation of some Yukawa couplings.

Our investigation shows how the variations of different couplings in the Standard Model may be compared. If the observational situation becomes clearer and at least one nonzero time variation is established, such methods may be used for new tests of the idea of grand unification.

\subsection*{Note added}
Shortly before the completion of this paper a new determination of the variation of $\mu$ appeared \cite{King:2008ud} reporting a reanalysis of spectra from the same two H$_2$ absorption systems as \cite{Reinhold:2006zn}, and adding one additional system at $z\simeq 2.8$. The results of the new analysis are not consistent with the previous claim indicating a nonzero variation, either considering all three systems or the two previously considered. The stringent null bound of the new analysis, $\Delta \mu/\mu = (2.6 \pm 3.0) \times 10^{-6}$, would disfavour all scenarios except those where the fractional variation of $\mu$ was of the same order as or smaller than that of $\alpha$. This would require us to approach the ``special'', apparently fine-tuned values of $\tilde{\gamma}$ discussed in Section~\ref{sec:Specials}, for which $\mu$ variation (and any deviation from the standard \lise\ abundance at BBN) are suppressed.

\subsection*{Acknowledgements}

We acknowledge useful discussions with M.~Pospelov, R.~Trotta, P.~Molaro, P.~Avelino, J.~Berengut and V.~Flambaum, and invaluable correspondence and discussions with M.~Murphy. T.\,D. is supported by the {\em Impuls- and Vernetzungsfond der Helmholtz-Gesellschaft}.

\appendix
\newcommand{\appsection}[1]{\let\oldthesection\thesection
  \renewcommand{\thesection}{Appendix \oldthesection}
  \section{#1}\let\thesection\oldthesection}

\appsection{Effect of ``varying constants'' at CMB and $\eta$} \label{app:CMBeta}

In our previous work on BBN we used the WMAP determination of the baryon number density parameter $\eta\equiv n_B/n_\gamma$ directly to reduce by one the number of unknown parameters. However, we should also consider the effect of possible variations of $G_k$ at the epoch of CMB decoupling. This question has distinct aspects: first, can the CMB alone or combined with various other cosmological observations give useful bounds on the values of fundamental parameters at this epoch? Second, how do the possible variations affect the determination of $\eta$? 

It would not be appropriate to give an extended discussion of CMB bounds on fundamental variations here; the subject has already been treated \cite{Trotta03} at length. Bounds tend to depend strongly on the values taken by cosmological parameters which are not at present well known through independent measurements: in other words there is considerable degeneracy. Fundamental parameters affecting the CMB are the proton and electron masses, the gravitational constant and the fine structure constant, as well as the mass of any dark matter particle present. In Planck units, these reduce to the particle masses and $\alpha$. The relevant cosmological parameters are the amplitude, spectral index (and possible running, {\em etc.})\ of primordial perturbations; the baryon, dark matter and dark energy (cosmological constant, {\em etc.})\ densities normalised to the critical density; the Hubble constant; and the reionization optical depth. Of these, the baryon density $\Omega_b h^2$ will vary linearly with the proton mass in Planck units, for a fixed baryon-to-photon ratio $\eta$. Conversely, given a measurement of $\Omega_b h^2$, the correct value of $\eta$ varies inversely with the proton mass. The conversion factor between $\Omega_b h^2$ and $\eta_{10}\equiv 10^{10} \eta$ is then $273.9 (m_p\sqrt{G})_{|0} (m_p\sqrt{G})^{-1} \simeq 273.9 (1-\Delta\ln(m_N/M_{\rm P})_{|\rm CMB} )$, where we approximate the proton and neutron masses by their average $m_N$.

If, therefore, we allow the proton mass (or the gravitational constant, in QCD units) to vary arbitrarily at the CMB epoch, $\eta$ is undetermined by WMAP and we must consider it as an extra free parameter or try to impose independent cosmological bounds. However, we impose that the size of variations away from the present value of $m_p/M_{\rm P}$ is a monotonically decreasing function of time: thus $\Delta\ln(m_N/M_{\rm P})_{|\rm CMB}\leq \Delta\ln(m_N/M_{\rm P})_{|\rm BBN}$. Hence we would have a self-consistent treatment of this parameter if the secondary discrepancies in primordial abundances due to an incorrectly estimated $\eta$ were smaller than the primary effect of varying $m_N/M_{\rm P}$ at BBN. The relevant results of our previous analysis
\begin{align}
 & \frac{\partial \ln ({\rm D/H}, Y_{\rm p}, \mbox{\lise/H}) }{\partial \ln (m_N/M_{\rm P})_{|\rm BBN}} &= &\ (1.88,0.72,-1.14), \nonumber \\
 \frac{\partial \ln ({\rm D/H}, Y_{\rm p}, \mbox{\lise/H}) }{\partial \ln \eta}
 = - &\frac{\partial \ln ({\rm D/H}, Y_{\rm p}, \mbox{\lise/H})}{\partial \ln (m_N/M_{\rm P})_{|\rm CMB}} &= &\ (-1.6, 0.04, 2.1)
\end{align}
are derived in QCD units where the strong coupling scale $\Lambda_c$ is constant, and where we neglect small contributions to the nucleon mass $m_N$ and take it proportional to $\Lambda_c$. The first relation, derived at a fixed value of $\eta=6.1\times 10^{-10}$ (WMAP3 \cite{WMAP3})\footnote{Updating to WMAP5 values does not lead to any significant change} led to the bound $-0.095\leq \Delta \ln (m_N/M_{\rm P})_{|\rm BBN} \leq +0.05$, where the main sensitivity to this variation is due to helium-4 ($Y_{\rm p}$). Since this abundance is insensitive to changes in $\eta$, we postulate also that $-0.095 \leq \Delta \ln (m_N/M_{\rm P})_{|\rm CMB} \leq 0.05$. 

%Thus we have $-0.05\leq \delta \ln \eta \leq 0.1$, where $\delta \ln\eta$ is the fractional discrepancy of the true value of $\eta$ at BBN compared to its value deduced from WMAP assuming the present value of $m_N/M_{\rm P}$. 

The resulting errors in the (standard) BBN abundances due to a possibly misestimated $\eta$ are then
\beq
	\delta \ln ({\rm D/H}, Y_{\rm p}, \mbox{\lise/H}) = (\{-0.15, 0.08 \}, \{-0.002, 0.004\}, \{-0.10, +0.20 \})
\eeq
to be compared with observational errors of 
\begin{gather}
	\sigma_{\rm D}/(\mbox{D/H}) \simeq 0.4/2.6 \simeq 0.15 \nonumber \\ 
	\sigma_{\rm 4He}/Y_{\rm p} \simeq 0.009/0.25 \simeq 0.04 \nonumber \\
	\sigma_{\rm 7Li}/(\mbox{\lise/H}) \simeq 0.5/4.5 \simeq 0.1,
\end{gather}
where we take the standard BBN \lise\ abundance $4.5\times 10^{-10}$ as central value. Hence the variation of $m_N/M_{\rm P}$ at the CMB epoch and consequent rescaling of $\eta$ may in principle have significant consequences for deuterium and lithium abundances in BBN. It may be appropriate to take $\Delta \ln (m_N/M_{\rm P})_{|\rm CMB}$ as an independent variable in the analysis of BBN variations.
The maximum effect due to rescaling of $\eta$ would occur when $\Delta \ln (m_N/M_{\rm P})_{|\rm CMB}$ = $\Delta \ln (m_N/M_{\rm P})_{|\rm BBN}$, giving a total sensitivity of
\beq
	\frac{\partial \ln ({\rm D/H}, Y_{\rm p}, \mbox{\lise/H})}{\partial \ln (m_N/M_{\rm P})_{|\rm BBN,CMB}} = (3.48,0.68,-3.24). \\
\eeq

\appsection{The \beei\ resonance} \label{app:Be8}

The \bese$ + n \rightarrow$\lise$ + p$ reaction is the main channel for destruction of \bese\ during BBN. If this reaction was not present, the final \lise\ abundance predicted by standard BBN would be considerably higher:
\[
\mbox{\lise /H} = 4.5 \times 10^{-10} \rightarrow\ \ \sim 14 \times 10^{-10}.
\]
The high cross section of this reaction is due to a strong \beei\ resonance which sits at about the energy of both %(the ground states of) 
\bese$+n$ and \lise$+p$ \cite{Ajzenberg-Selove}. For the reaction to continue to operate efficiently, it is important that the resonance remains near these \bese\ / \lise\ energy levels. We will argue here that, given the size of coupling variations relevant for our paper, this is indeed the case.

In \cite{DSW07} we estimated the dependence of nuclear binding energies on the pion mass by
\beq \label{dBdmpi}
 \frac{\partial B_i}{\partial m_\pi} = f_i (A_i-1) \frac{B_{\rm D}}{m_\pi} r \simeq
 -0.13 f_i (A_i-1),
\eeq
taking $r\simeq-8$. The constants $f_i$ are expected to be of order unity, but will differ between light nuclei due to peculiarities of the shell structure. Our normalization corresponds to $f_{\rm D}=1$. We are then concerned with the relative changes of the \bese\ and \lise\ binding energies and the energy of the \beei\ resonance, whose dependence we will estimate in an analogous way with a constant of proportionality $f'_8$. Then
\bea 
	\Delta B_{\rm 7Be} &=& -9.1\, \mev \times 6 f_{\rm 7Be}\, \Delta \ln \hat{m},  \nonumber \\
	\Delta E_{\rm 8Be^*} &=& -9.1\, \mev \times 7 f'_{8}\, \Delta \ln \hat{m}, \nonumber \\
	\Delta B_{\rm 7Li} &=& -9.1\, \mev \times 6 f_{\rm 7Li}\, \Delta \ln \hat{m},
\eea
recalling that $m_\pi\propto \hat{m}^{1/2}$. In \cite{Adahchour03} the sum of the neutron and proton widths of the \beei\ resonance is given as approximately $1.6\,\mev$ %(Ajzenberg-Selove?) 
thus for the \beei-destroying reaction to remain effective we require at least
\beq
|9.1\,\mev (7 f'_8 - 6 f_7) \Delta \ln \hat{m} | < 1.6\, \mev,
\eeq
where $f_7$ may correspond to either \bese\ or \lise. If we take all $f_i = 1$ this condition becomes $|\Delta \ln \hat{m}| < 0.18$, easily satisfied by the range of variations that we consider ($\Delta \ln \hat{m}$ was bounded at about $1.5\%$). However, this would imply a substantial cancellation between the variations of $A=7$ and $A=8$ states, which may not occur for the true values of $f_i$. There may be less cancellation, for example if $f'_8 = 2,$ $f_7 = 0.5$ we obtain $|\Delta \ln \hat{m}| < 0.016$, which is still fulfilled in the unified scenarios we consider where the variation of $\hat{m}$ is around $1\%$.


\begin{thebibliography}{99} 


\bibitem{DSW07}
  T.~Dent, S.~Stern and C.~Wetterich,
  %``Primordial nucleosynthesis as a probe of fundamental physics parameters,''
  Phys.\ Rev.\  D {\bf 76}, 063513 (2007)
  [0705.0696 [astro-ph]].
  %%CITATION = PHRVA,D76,063513;%%

\bibitem{CocNunes06}
  A.~Coc, N.\,J.~Nunes, K.\,A.~Olive, J.-P.~Uzan and E.~Vangioni,
  %``Coupled Variations of Fundamental Couplings and Primordial
  %Nucleosynthesis,''
    Phys.\ Rev.\  D {\bf 76} (2007) 023511
  [astro-ph/0610733].
  %%CITATION = PHRVA,D76,023511;%%

\bibitem{Wetterich:1987fk}
  C.~Wetterich,
  %``COSMOLOGIES WITH VARIABLE NEWTON'S 'CONSTANT',''
  Nucl.\ Phys.\  B {\bf 302} (1988) 645.
  %%CITATION = NUPHA,B302,645;%%

\bibitem{DvaliZ}
  G.~R.~Dvali and M.~Zaldarriaga,
  %``Changing alpha with time: Implications for fifth-force-type experiments
  %and quintessence,''
  Phys.\ Rev.\ Lett.\  {\bf 88} (2002) 091303.
%  [hep-ph/0108217].
  %%CITATION = PRLTA,88,091303;%%

\bibitem{Wetterich:2002ic}
  C.~Wetterich,
  %``Probing quintessence with time variation of couplings,''
  JCAP {\bf 0310} (2003) 002
  [hep-ph/0203266].
  %%CITATION = JCAPA,0310,002;%%

\bibitem{Nunes:2003ff}
  N.~J.~Nunes and J.~E.~Lidsey,
  %``Reconstructing the dark energy equation of state with varying alpha,''
  Phys.\ Rev.\  D {\bf 69} (2004) 123511
%  [astro-ph/0310882].
  %%CITATION = PHRVA,D69,123511;%%

\bibitem{Part2}
  T.~Dent, S.~Stern and C.~Wetterich, ``Time variation of fundamental couplings and dynamical dark energy,'' eprint arXiv:0809.4628.

\bibitem{Will}
  C.~M.~Will,
  %``The confrontation between general relativity and experiment,''
  Living Rev.\ Rel.\  {\bf 9} (2005) 3
  [gr-qc/0510072].
  %%CITATION = 00222,9,3;%%

\bibitem{LiChu05}
  B.~Li and M.~C.~Chu,
  %``Big bang nucleosynthesis with an evolving radion in the brane world
  %scenario,''
  Phys.\ Rev.\  D {\bf 73} (2006) 023509;
%  [astro-ph/0511642];
  %%CITATION = PHRVA,D73,023509;%%
  %``Big bang nucleosynthesis constraints on universal extra dimensions and
  %varying fundamental constants,''
  Phys.\ Rev.\  D {\bf 73} (2006) 025004.
%  [hep-ph/0511013].
  %%CITATION = PHRVA,D73,025004;%%

\bibitem{Chamoun05}
  N.~Chamoun, S.\,J.~Landau, M.\,E.~Mosquera and H.~Vucetich,
  %``Helium and deuterium abundances as a test for the time variation of the
  %fine structure constant and the Higgs vacuum expectation value,''
  J.\ Phys.\ G {\bf 34} (2007) 163.
%  [astro-ph/0508378].
  %%CITATION = JPHGB,G34,163;%%

\bibitem{Landau04}
  S.\,J.~Landau, M.\,E.~Mosquera and H.~Vucetich,
  %``Primordial nucleosynthesis with varying fundamental constants: A
  %semi-analytical approach,''
  Astrophys.\ J.\  {\bf 637} (2006) 38.
%  [astro-ph/0411150].
  %%CITATION = ASJOA,637,38;%%

\bibitem{MSW}
  C.\,M.~M{\"u}ller, G.~Sch{\"a}fer and C.~Wetterich,
  %``Nucleosynthesis and the variation of fundamental couplings,''
  Phys.\ Rev.\  D {\bf 70} (2004) 083504
  [astro-ph/0405373].
  %%CITATION = PHRVA,D70,083504;%%

\bibitem{Dmitriev03}
  V.~F.~Dmitriev, V.~V.~Flambaum and J.~K.~Webb,
  %``Cosmological variation of deuteron binding energy, strong interaction  and
  %quark masses from big bang nucleosynthesis,''
  Phys.\ Rev.\  D {\bf 69} (2004) 063506.
%  [astro-ph/0310892].
  %%CITATION = PHRVA,D69,063506;%%

\bibitem{ScherrerGN}
  R.\,J.~Scherrer,
  %``The uncertainty in Newton's constant and precision predictions of the
  %primordial helium abundance,''
  Phys.\ Rev.\  D {\bf 69} (2004) 107302.
%  [astro-ph/0310699].
  %%CITATION = PHRVA,D69,107302;%%

\bibitem{KnellerLambda}
  J.\,P.~Kneller and G.\,C.~McLaughlin,
  %``BBN and Lambda(QCD),''
  Phys.\ Rev.\ D {\bf 68} (2003) 103508.
%  [nucl-th/0305017].
  %%CITATION = PHRVA,D68,103508;%%

\bibitem{YooScherrer}
  J.\,J.~Yoo and R.\,J.~Scherrer,
  %``Big bang nucleosynthesis and cosmic microwave background constraints on the
  %time variation of the Higgs vacuum expectation value,''
  Phys.\ Rev.\  D {\bf 67} (2003) 043517.
%  [astro-ph/0211545].
  %%CITATION = PHRVA,D67,043517;%%

\bibitem{NollettLopez}
  K.\,M.~Nollett and R.\,E.~Lopez,
  %``Primordial nucleosynthesis with a varying fine structure constant: An
  %improved estimate,''
  Phys.\ Rev.\  D {\bf 66} (2002) 063507.
%  [astro-ph/0204325].
  %%CITATION = PHRVA,D66,063507;%%

\bibitem{DentF}
  T.~Dent and M.~Fairbairn,
  %``Time-varying coupling strengths, nuclear forces and unification,''
  Nucl.\ Phys.\  B {\bf 653} (2003) 256.
%  [arXiv:hep-ph/0112279].
  %%CITATION = NUPHA,B653,256;%%

\bibitem{CampbellPrimordialSoup}
  B.\,A.~Campbell and K.\,A.~Olive,
  %``Nucleosynthesis and the time dependence of fundamental couplings,''
  Phys.\ Lett.\  B {\bf 345} (1995) 429.
%  [arXiv:hep-ph/9411272].
  %%CITATION = PHLTA,B345,429;%%

\bibitem{Uzanreview}
  J.-P.~Uzan,
  %``The fundamental constants and their variation: Observational status and
  %theoretical motivations,''
  Rev.\ Mod.\ Phys.\  {\bf 75} (2003) 403.
%  [arXiv:hep-ph/0205340].
  %%CITATION = RMPHA,75,403;%%

\bibitem{WMAP5}
  J.~Dunkley {\it et al.}  [WMAP Collaboration],
  %``Five-Year Wilkinson Microwave Anisotropy Probe (WMAP) Observations:
  %Likelihoods and Parameters from the WMAP data,''
  arXiv:0803.0586 [astro-ph];
  %%CITATION = ARXIV:0803.0586;
  G.~Hinshaw {\it et al.}  [WMAP Collaboration],
  %``Five-Year Wilkinson Microwave Anisotropy Probe (WMAP\altaffilmark 1 )
  %Observations:Data Processing, Sky Maps, \& Basic Results,''
  arXiv:0803.0732 [astro-ph].

\bibitem{Fiorentini:1998fv}
  G.~Fiorentini, E.~Lisi, S.~Sarkar and F.~L.~Villante,
  %``Quantifying uncertainties in primordial nucleosynthesis without Monte
  %Carlo simulations,''
  Phys.\ Rev.\  D {\bf 58}, 063506 (1998).
%  [astro-ph/9803177].
  %%CITATION = PHRVA,D58,063506;%%

\bibitem{Korn06}
  A.\,J.~Korn {\it et al.},
  %``A probable stellar solution to the cosmological lithium discrepancy,''
  Nature {\bf 442} (2006) 657
  [astro-ph/0608201].
  %%CITATION = NATUA,442,657;%%

\bibitem{Martins:2003pe}
  C.~J.~A.~Martins, A.~Melchiorri, G.~Rocha, R.~Trotta, P.~P.~Avelino and P.~Viana,
  %``WMAP Constraints on varying $\alpha$ and the Promise of Reionization,''
  Phys.\ Lett.\  B {\bf 585}, 29 (2004).
%  [astro-ph/0302295].
  %%CITATION = PHLTA,B585,29;%%

\bibitem{Trotta03}
  G.~Rocha {\it et al.}, 
%R.~Trotta, C.\,J.\,A.~Martins, A.~Melchiorri, P.\,P.~Avelino, R.~Bean and P.\,T.\,P.~Viana,
  %``Measuring $\alpha$ in the Early Universe: CMB Polarization, Reionization
  %and the Fisher Matrix Analysis,''
  Mon.\ Not.\ Roy.\ Astron.\ Soc.\  {\bf 352} (2004) 20.
%  [astro-ph/0309211].
  %%CITATION = MNRAA,352,20;%%

\bibitem{Chan:2007fe}
  K.~C.~Chan and M.~C.~Chu,
  %``Constraining the Variation of G by Cosmic Microwave Background
  %Anisotropies,''
  Phys.\ Rev.\  D {\bf 75}, 083521 (2007).
%  [astro-ph/0611851].
  %%CITATION = PHRVA,D75,083521;%%

\bibitem{ZahnZaldarriaga02}
  O.~Zahn and M.~Zaldarriaga,
  %``Probing the Friedmann equation during recombination with future CMB
  %experiments,''
  Phys.\ Rev.\  D {\bf 67} (2003) 063002.
%  [astro-ph/0212360].
  %%CITATION = PHRVA,D67,063002;%%

\bibitem{Murphy:2003mi}
  M.\,T.~Murphy, V.\,V.~Flambaum, J.\,K.~Webb, V.\,V.~Dzuba, J.\,X.~Prochaska and A.\,M.~Wolfe,
  %``Constraining variations in the fine-structure constant, quark masses  and
  %the strong interaction,''
  Lect.\ Notes Phys.\  {\bf 648}, 131 (2004)
  [astro-ph/0310318].
  %%CITATION = LNPHA,648,131;%%

\bibitem{Murphy0306}
  M.\,T.~Murphy, J.\,K.~Webb and V.\,V.~Flambaum,
  %``Further evidence for a variable fine-structure constant from Keck/HIRES
  %QSO absorption spectra,''
  Mon.\ Not.\ Roy.\ Astron.\ Soc.\  {\bf 345} (2003) 609.
%  [astro-ph/0306483].
  %%CITATION = MNRAA,345,609;%%

\bibitem{Murphyprivate}
  M.\,T.~Murphy, private communication.

\bibitem{Fujii:2007zg}
  Y.~Fujii,
  %``Possible time-variability of the fine-structure constant expected from the
  %accelerating universe,''
  Phys.\ Lett.\  B {\bf 660} (2008) 87
  [arXiv:0709.2211 [astro-ph]].
  %%CITATION = PHLTA,B660,87;%%

\bibitem{Levshakov:2007da}
  S.\,A.~Levshakov {\it et al.},
  %``A new measure of Delta alpha/alpha at redshift z = 1.84 from very high
  %resolution spectra of Q1101-264,''
  astro-ph/0703042.
  %%CITATION = ASTRO-PH/0703042;%%

\bibitem{Srianand04}
  R.~Srianand, H.~Chand, P.~Petitjean and B.~Aracil,
  %``Limits on the time variation of the electromagnetic fine-structure
  %constant in the low energy limit from absorption lines in the spectra  of
  %distant quasars,''
  Phys.\ Rev.\ Lett.\  {\bf 92} (2004) 121302.
%  [astro-ph/0402177].
  %%CITATION = PRLTA,92,121302;%%

\bibitem{Levshakov:2005ab}
  S.~A.~Levshakov {\it et al.}, 
%M.~Centurion, P.~Molaro, S.~D'Odorico, D.~Reimers, R.~Quast and M.~Pollmann,
  %``Most precise single redshift bound to Delta alpha/alpha,''
  Astron.\ Astrophys. {\bf 449} (2006) 879.
%  [astro-ph/0511765].
  %%CITATION = ASTRO-PH/0511765;%%

\bibitem{Murphycrit}
  M.~T.~Murphy, J.~K.~Webb and V.~V.~Flambaum,
  %``Revision of VLT/UVES constraints on a varying fine-structure constant,''
  Mon.\ Not.\ Roy.\ Astron.\ Soc.\  {\bf 384} (2008) 1053
  [astro-ph/0612407].

\bibitem{Molaro:2007kp}
  P.~Molaro, D.~Reimers, I.~I.~Agafonova and S.~A.~Levshakov,
  %``Bounds on the fine structure constant variability from FeII absorption
  %lines in QSO spectra,''
  0712.4380 [astro-ph].
  %%CITATION = ARXIV:0712.4380;%%

\bibitem{Reinhold:2006zn}
  E.~Reinhold, R.~Buning, U.~Hollenstein, A.~Ivanchik, P.~Petitjean and W.~Ubachs,
  %``Indication of a Cosmological Variation of the Proton - Electron Mass Ratio
  %Based on Laboratory Measurement and Reanalysis of H(2) Spectra,''
  Phys.\ Rev.\ Lett.\  {\bf 96}, 151101 (2006).
  %%CITATION = PRLTA,96,151101;%%

\bibitem{Wendt08}
  M.~Wendt and D.~Reimers,
  %``Variability of the proton-to-electron mass ratio on cosmological scales,''
  0802.1160 [astro-ph].
  %%CITATION = ARXIV:0802.1160;%%

\bibitem{FlambaumNH3}
  V.~V.~Flambaum and M.~G.~Kozlov,
  %``Enhanced sensitivity to time-variation of m(p)/m(e) in the inversion
  %spectrum of ammonia,''
  Phys.\ Rev.\ Lett.\ {\bf 98}, 240801 (2007)
  [0704.2301 [astro-ph]].
  %%CITATION = ARXIV:0704.2301;%%

\bibitem{MurphyNH3}
  M.~T.~Murphy, V.~V.~Flambaum, S.~Muller and C.~Henkel,
  %``Strong Limit on a Variable Proton-to-Electron Mass Ratio from Molecules in
  %the Distant Universe,''
  Science {\bf 320}, 1611 (2008)
  [0806.3081 [astro-ph]].
  %%CITATION = ARXIV:0806.3081;%%

\bibitem{Murphy:2001nu}
  M.~T.~Murphy {\it et al.}, 
%J.~K.~Webb, V.~V.~Flambaum, M.~J.~Drinkwater, F.~Combes and T.~Wiklind,
  %``Improved constraints on possible variation of physical constants from H I
  %21cm and molecular QSO absorption lines,''
  Mon.\ Not.\ Roy.\ Astron.\ Soc.\  {\bf 327}, 1244 (2001).
%  [astro-ph/0101519].
  %%CITATION = MNRAA,327,1244;%%

\bibitem{Tzanavaris:2006uf}
  P.~Tzanavaris, M.~T.~Murphy, J.~K.~Webb, V.~V.~Flambaum and S.~J.~Curran,
  %``Probing variations in fundamental constants with radio and optical quasar
  %absorption-line observations,''
  Mon.\ Not.\ Roy.\ Astron.\ Soc.\  {\bf 374}, 634 (2007)
  [astro-ph/0610326].
  %%CITATION = MNRAA,374,634;%%

\bibitem{Kanekar:2005xy}
  N.~Kanekar {\it et al.},
  %``Constraints on changes in fundamental constants from a cosmologically
  %distant OH absorber / emitter,''
  Phys.\ Rev.\ Lett.\  {\bf 95} (2005) 261301.
%  [astro-ph/0510760].
  %%CITATION = PRLTA,95,261301;%%

\bibitem{Levshakov:2007tn}
  S.\,A.~Levshakov, D.~Reimers, M.\,G.~Kozlov, S.\,G.~Porsev and P.~Molaro,
  %``A new approach for testing variations of fundamental constants over cosmic
  %epochs using FIR fine-structure lines,''
  arXiv:0712.2890 [astro-ph].
  %%CITATION = ARXIV:0712.2890;%%

\bibitem{Petrov:2005pu}
  Yu.~V.~Petrov, A.~I.~Nazarov, M.~S.~Onegin, V.~Y.~Petrov and E.~G.~Sakhnovsky,
  %``Natural nuclear reactor Oklo and variation of fundamental constants. I:
  %Computation of neutronic of fresh core,''
  Phys.\ Rev.\  C {\bf 74}, 064610 (2006).
%  [hep-ph/0506186].
  %%CITATION = PHRVA,C74,064610;%%

\bibitem{Gould}
  C.~R.~Gould, E.~I.~Sharapov and S.~K.~Lamoreaux,
  %``Time-variability of alpha from realistic models of Oklo reactors,''
  Phys.\ Rev.\  C {\bf 74} (2006) 024607.
%  [nucl-ex/0701019].
  %%CITATION = PHRVA,C74,024607;%%

\bibitem{Olive:2002tz}
  K.~A.~Olive, M.~Pospelov, Y.~Z.~Qian, A.~Coc, M.~Casse and E.~Vangioni-Flam,
  %``Constraints on the variations of the fundamental couplings,''
  Phys.\ Rev.\  D {\bf 66}, 045022 (2002).
%  [hep-ph/0205269].
  %%CITATION = PHRVA,D66,045022;%%

\bibitem{Flambaum:2002wq}
  V.~V.~Flambaum and E.~V.~Shuryak,
  %``Dependence of hadronic properties on quark masses and constraints on  their
  %cosmological variation,''
  Phys.\ Rev.\  D {\bf 67} (2003) 083507.
%  [hep-ph/0212403].
  %%CITATION = PHRVA,D67,083507;%%

\bibitem{Olive:2003sq}
  K.~A.~Olive {\em et al.}, 
%M.~Pospelov, Y.~Z.~Qian, G.~Manhes, E.~Vangioni-Flam, A.~Coc and M.~Casse,
  %``A re-examination of the Re-187 bound on the variation of fundamental
  %couplings,''
  Phys.\ Rev.\  D {\bf 69}, 027701 (2004)
% [astro-ph/0309252].
  %%CITATION = PHRVA,D69,027701;%%

\bibitem{SisternaV}
%\bibitem{Sisterna:1991zt}
  P.~Sisterna and H.~Vucetich,
  %``Time variation of fundamental constants: 2: Quark masses as time dependent
  %parameters,''
  Phys.\ Rev.\  D {\bf 44} (1991) 3096,
  %%CITATION = PHRVA,D44,3096;%%
%\cite{Sisterna:1990et}
%\bibitem{Sisterna:1990et}
%  P.~Sisterna and H.~Vucetich,
  %``Time variation of fundamental constants: Bounds from geophysical and
  %astronomical data,''
  Phys.\ Rev.\  D {\bf 41} (1990) 1034. 
  %%CITATION = PHRVA,D41,1034;%%

\bibitem{FujiiMeteorites}
  Y.~Fujii and A.~Iwamoto,
  %``How strongly does dating meteorites constrain the time-dependence of  the
  %fine-structure constant?,''
  Mod.\ Phys.\ Lett.\  A {\bf 20} (2005) 2417,
  %``Re/Os constraint on the time-variability of the fine-structure constant,''
  Phys.\ Rev.\ Lett.\  {\bf 91} (2003) 261101.
%  [hep-ph/0309087].
  %%CITATION = PRLTA,91,261101;%%

\bibitem{DentEotvos}
  T.~Dent,
  %``E{\" o}tv{\" o}s bounds on couplings of fundamental parameters to gravity,''
  Phys.\ Rev.\ Lett\ {\bf 101} (2008) 041102
  [0805.0318 [hep-ph]].
  %%CITATION = ARXIV:0805.0318;%%

\bibitem{Hellings83}
  R.~W.~Hellings {\it et al.},
  Phys.\ Rev.\ Lett.\ 51 (1983) 1609.

\bibitem{Williams04}
  J.~G.~Williams, S.~G.~Turyshev and D.~H.~Boggs,
  %``Progress in lunar laser ranging tests of relativistic gravity,''
  Phys.\ Rev.\ Lett.\  {\bf 93} (2004) 261101.
%  [arXiv:gr-qc/0411113].
  %%CITATION = PRLTA,93,261101;%%

\bibitem{Damour88}
  T.~Damour, G.\,W.~Gibbons, J.\,H.~Taylor,
  %``Limits on the Variability of $G$ Using Binary-Pulsar Data,''
  Phys.\ Rev.\ Lett.\  {\bf 61} (1988) 1151.

\bibitem{Deller:2008jx}
  A.~T.~Deller, J.~P.~W.~Verbiest, S.~J.~Tingay and M.~Bailes,
  %``Extremely high precision VLBI astrometry of PSR J0437-4715 and implications
  %for theories of gravity,''
  arXiv:0808.1594 [astro-ph].
  %%CITATION = ARXIV:0808.1594;%% 

\bibitem{Krauss:1997}
  D.~B.~Guenther, L.~M.~Krauss and P.~Demarque,
  %``Testing the constancy of the gravitational constant using helioseismology,''
  Astrophys.\ J.\  {\bf 498} (1998) 871.

\bibitem{Jofre:2006ug}
  P.~Jofre, A.~Reisenegger and R.~Fernandez,
  %``Constraining a possible time-variation of the gravitational constant
  %through gravitochemical heating of neutron stars,''
  Phys.\ Rev.\ Lett.\  {\bf 97} (2006) 131102.
%  [astro-ph/0606708].
  %%CITATION = PRLTA,97,131102;%%

\bibitem{Benvenuto:2004bs}
  O.~G.~Benvenuto, E.~Garcia-Berro and J.~Isern,
  %``Asteroseismological bound on G/G from pulsating white dwarfs,''
  Phys.\ Rev.\  D {\bf 69} (2004) 082002.
  %%CITATION = PHRVA,D69,082002;%%

\bibitem{Thorsett:1996fr}
  S.~E.~Thorsett,
  %``The Gravitational Constant, the Chandrasekhar Limit, and Neutron Star
  %Masses,''
  Phys.\ Rev.\ Lett.\  {\bf 77}, 1432 (1996).
%  [astro-ph/9607003].
  %%CITATION = PRLTA,77,1432;%%

\bibitem{Peik:2006xy}
  E.~Peik, B.~Lipphardt, H.~Schnatz, C.~Tamm, S.~Weyers and R.~Wynands,
  %``Laboratory limits on temporal variations of fundamental constants: An
  %update,''
  physics/0611088.
  %%CITATION = PHYSICS/0611088;%%

\bibitem{Blatt:2008su}
  S.~Blatt {\it et al.},
  %``New Limits on Coupling of Fundamental Constants to Gravity Using $^{87}$Sr
  %Optical Lattice Clocks,''
  Phys.\ Rev.\ Lett.\  {\bf 100} (2008) 140801
  [0801.1874 [physics.atom-ph]].
  %%CITATION = PRLTA,100,140801;%%

\bibitem{Fortier:2007jf}
  T.~M.~Fortier {\it et al.},
  %``Precision Atomic Spectroscopy For Improved Limits On Variation Of The Fine
  %Structure Constant And Local Position Invariance,''
  Phys.\ Rev.\ Lett.\  {\bf 98}, 070801 (2007).
  %%CITATION = PRLTA,98,070801;%%

\bibitem{Rosenband}
  T.~Rosenband {\it et al.}, 
  %``Frequency Ratio of Al+ and Hg+ Single-Ion Optical Clocks; Metrology at the 
  %17th Decimal Place,''
  Science {\bf 319} (2008), 1808.

\bibitem{Dent03}
  T.~Dent,
  ``Varying alpha, thresholds and extra dimensions,''
  hep-ph/0305026.
  %%CITATION = HEP-PH/0305026;%%

\bibitem{CalmetLangacker}
  X.~Calmet and H.~Fritzsch,
  %``The cosmological evolution of the nucleon mass and the electroweak
  %coupling constants,''
  Eur.\ Phys.\ J.\  C {\bf 24} (2002) 639;
% [arXiv:hep-ph/0112110].
  %%CITATION = EPHJA,C24,639;%%
  P.~Langacker, G.~Segre and M.\,~Strassler,
  %``Implications of gauge unification for time variation of the fine  structure
  %constant,''
  Phys.\ Lett.\  B {\bf 528} (2002) 121.
%  [arXiv:hep-ph/0112233].
  %%CITATION = PHLTA,B528,121;%%

\bibitem{King:2008ud}
  J.~A.~King, J.~K.~Webb, M.~T.~Murphy and R.~F.~Carswell,
  %``Stringent null constraint on cosmological evolution of the
  %proton-to-electron mass ratio,''
  arXiv:0807.4366 [astro-ph].
  %%CITATION = ARXIV:0807.4366;%%

\bibitem{WMAP3}
  D.~N.~Spergel {\it et al.}  [WMAP Collaboration],
  %``Wilkinson Microwave Anisotropy Probe (WMAP) three year results:
  %Implications for cosmology,''
  Astrophys.\ J.\ Suppl.\  {\bf 170}, 377 (2007).
%  [astro-ph/0603449].
  %%CITATION = APJSA,170,377;%%

%\cite{AjzenbergSelove:1988ec}
\bibitem{Ajzenberg-Selove}
  F.~Ajzenberg-Selove,
  %``Energy levels of light nuclei A = 5-10,''
  Nucl.\ Phys.\  A {\bf 490} (1988) 1.
  %%CITATION = NUPHA,A490,1;%%

\bibitem{Adahchour03}
  A.~Adahchour and P.~Descouvemont,
  %``R-matrix analysis of the 3He(n,p)3H and 7Be(n,p)7Li reactions,''
  J.\ Phys. G {\bf 29} (2003) 395. 

\end{thebibliography}
\end{document}